\documentstyle[floats,12pt,epsf,aps,preprint]{revtex}
 
 \newcommand{\be}{\begin{eqnarray}}
 \newcommand{\ee}{\end{eqnarray}}
 \newcommand{\non}{\nonumber\\}
 \newcommand{\scrpt}[1]{{\hbox{\scriptsize #1}}}
 \newcommand{\inline}[1]{\noalign{\hbox{#1}}}
 \newcommand{\bgamma}{\hbox{\boldmath{$\gamma$}}} 
 \newcommand{\dom}[1]{{d #1 \over 2\pi}}
 \newcommand{\dum}{{\vphantom{x}}} 

\begin{document}
 \draft
 \title{
 \begin{flushright}
 {\normalsize NUC-MINN-97/13-T\\}
 \end{flushright}
 \vspace{1.5cm}
 {\bf Multiple Scattering Expansion of the Self-Energy at Finite Temperature}
 }

 \author{Sangyong Jeon\footnote{jeon@nucth1.spa.umn.edu} 
 \ and Paul J. Ellis\footnote{ellis@physics.spa.umn.edu}\\
 \it School of Physics and Astronomy\\
 \it University of Minnesota\\
 \it Minneapolis, MN 55455}
 

 \maketitle

 \begin{abstract}
 An often used rule that the thermal correction to the 
 self-energy is the thermal phase-space times 
 the forward scattering amplitude from target particles is shown to be
 the leading term in an exact multiple scattering expansion. 
 Starting from imaginary-time finite-temperature field theory, 
 a rigorous expansion for the retarded self-energy 
 is derived. The relationship to the thermodynamic potential is 
 briefly discussed.
 \end{abstract}
 \thispagestyle{empty}

 \newpage
 \setcounter{page}{1} 
 \section{Introduction}
 \label{sec:intro}

 At a temperature much lower than the particle mass, it is physically
 reasonable to expect that the leading order thermal correction 
 to the physical self-energy be given by 
 the forward scattering amplitude integrated over the 
 thermal phase-space of the lighter particle.
 For instance, in the literature, 
 the thermal self-energy of a particle 
 is often written as \cite{Shuryak,Ashida,Zahed}
 \be
 \Sigma_T(p) 
 =
 -\int {d^3 k\over (2\pi)^3 2 E_\scrpt{light}}\,
 n(E_\scrpt{light})\,
 {\cal T}(k+p \to k+p)
 \;,\label{eq:shuryak}
 \ee
 where 
 $E_\scrpt{light} = \sqrt{{\bf k}^2 + m_\scrpt{light}^2}$ is the energy of
 the lighter particle with mass $m_\scrpt{light}$,
 $n(E_\scrpt{light})$ is a Bose or Fermi distribution function, and
 ${\cal T}(k+p \to k+p)$ is the forward scattering amplitude
 related to the usual $f$ by   
 \be
 {\cal T}(k+p \to k+p)
 =
 8\pi \sqrt{s}\, f(k,p) 
 \;.
 \ee
 The minus sign on the right hand side of Eq.~(\ref{eq:shuryak})
 stems from the fact that the
 scattering amplitude is defined to be $-i$ times an
 $N$-point correlation
 function whereas the self-energy is defined to be $i$ times a 2-point
 correlation function.
 
 A merit of this method is that 
 one can simply take the scattering amplitudes from experiment.
 In reactions involving strong couplings, this may be the only reliable
 way of calculating the thermal correction to the real part of the
 self-energy \cite{Smilga}.  
 As the density and temperature become higher,
 Eq.~(\ref{eq:shuryak}) needs higher order corrections.
 In this paper, we start with the imaginary-time formulation of
 relativistic finite temperature field theory and  
 obtain an exact expression for the thermal self-energy.

 There are a variety of finite temperature
 field theory methods used in literature. 
 The real-time method was first investigated by
 Schwinger and Keldysh and later in a slightly different and modern
 form by many people \cite{Semenoff,Niemi,Kobes,Kobes90,Evans}.
 The imaginary-time method was introduced by Matsubara \cite{Kapusta}
 and through analytic
 continuation can be related to the real-time method \cite{Fetter}. 
 Of course any
 equilibrium Green function can be calculated using any of the methods,
 given the analytic relations between them.  

 There are, however, situations where one particular method is more
 economical.  For instance, if one would like to study 
 time-ordered real-time correlation functions, the real-time method
 is very much the natural choice.  
 If one would like to study a response function (a retarded function),
 the most economical way would be to compute the imaginary-time 
 correlation function with Matsubara frequencies and then analytically 
 continue the result.  
 
 In this paper we are interested in the retarded correlation functions.  
 We would like to use the imaginary-time method.  
 However, since the imaginary time is
 a fictitious parameter introduced to deal with the
 temperature, it is not easy to gain physical insight by just looking 
 at the end result of the imaginary time calculation.    
 A way to remedy this difficulty was developed by one of the present
 authors \cite{jeon} based on earlier works 
 \cite{Mills,Baym,Weldon}.
 There, a diagrammatic method to calculate the spectral density of
 two-point functions in a scalar theory was presented, starting from
 the imaginary-time formulation of finite temperature field theory. 
 Subsequently it was used to calculate the leading order hydrodynamic
 coefficients in a scalar theory \cite{jeon2}.
 In this paper the above work is
 extended to the calculation of $N$-point retarded functions.

 The analysis in this paper is based on the fact that
 applying the analytic continuation 
 \be
 i\nu \to k^0 + i\epsilon
 \label{eq:analytic_cont}
 \ee
 to each of the independent external Matsubara frequencies
 produces the (Fourier transformed) retarded correlation function 
 \cite{Evans,Baier}.
 When the chemical potentials all vanish, this operation produces the 
 physical retarded function.  However, with non-zero chemical potentials 
 one must be more careful. 

 In the imaginary-time formalism, finite density is dealt with 
 by using an effective Hamiltonian
 \be
 \hat{K} \equiv \hat{H} - \sum_{a} \mu_a \hat{Q}_a \;,
 \ee
 where $\mu_a$ is the chemical potential associated with each conserved
 charge $\hat{Q}_a$. 
 Hence, the retarded function produced by using
 Eq.~(\ref{eq:analytic_cont}) corresponds to the retarded correlation 
 function of 
 $\hat{\varphi}_K(t) \equiv e^{i\hat{K}t}\hat{\varphi}e^{-i\hat{K}t}$. 
 But the physical retarded function 
 is a correlation function of 
 $\hat{\varphi}_H(t) \equiv e^{i\hat{H}t}\hat{\varphi}e^{-i\hat{H}t}$, 
 not $\hat{\varphi}_K(t)$.
 Fortunately, there is a simple relation between the two.  
 In most cases the fields of interest are charge eigenstates. Then
 \be
 \hat{\varphi}_K(t) = e^{i\mu_\varphi t} \hat{\varphi}_H(t) \;,
 \ee
 where
 \be
 \mu_\varphi = \sum_{a}\mu_a q_a
 \ee
 is the total chemical potential associated with the field $\varphi$
 carrying conserved charges $q_a$.
 In momentum space, the two retarded functions are related by 
 \cite{Landsman}
 \be
 G_{\varphi_H}^\scrpt{Ret}(\{ k \})
 =
 G_{\varphi_K}^\scrpt{Ret}(\{ k - \mu_\varphi \})
 \;,\label{eq:ret_shift}
 \ee
 where we used a shorthand notation, 
 $k - \mu_\varphi \equiv (k^0 - \mu_\varphi, {\bf k})$.
 As is shown in Sec.~\ref{sec:retarded}, this shift also has the
 effect of separating the purely dynamic part of the theory and the
 purely statistical part.

 This paper is organized as follows.  In Sec.~\ref{sec:ret_func}
 a brief derivation is given for the diagrammatic rules 
 to compute an arbitrary retarded function.
 Section~\ref{sec:self_energy} presents a derivation of the self-energy 
 formula using the results from Sec.~\ref{sec:ret_func}. 
 As an example, the self-energy of an electron at the one-loop level
 is worked out in Sec.~\ref{sec:elec_self}.
 Finally, we conclude in Sec.~\ref{sec:concl}.
 Appendices~\ref{app:eucl} and \ref{app:real} list some relevant facts
 about the finite temperature propagators used in this paper. 
 Appendix~\ref{app:derivative} deals with derivative couplings.
 Appendix~\ref{app:omega} amplifies the discussion of 
 the thermodynamic potential in Sec.~\ref{sec:self_energy}.

 \section{$N$-point Retarded Correlation Functions}
 \label{sec:ret_func}

 In this section, as a prelude to the self-energy calculation,
 diagrammatic rules to calculate an arbitrary 
 $N$-point retarded function are derived starting 
 from imaginary-time finite temperature field theory.
 The derivation here closely follows Ref.~\cite{jeon}.  
 Details omitted for the sake of brevity and readability can be
 found in that reference and references therein. 
 The rules derived here coincide with the rules obtained by
 Evans~\cite{Evans} who used the real-time method 
 and the analytic relations between $N$-point functions. 
 For simplicity, only theories involving 
 no derivative couplings are considered here.
 However, the argument given here can be straightforwardly generalized
 to theories with derivative couplings, such as chiral
 perturbation theory, as indicated in
 Appendix~\ref{app:derivative}. 

 \subsection{Euclidean Correlation Functions}
 
 It is known that performing Matsubara
 frequency sums in an Euclidean Feynman diagram produces terms
 corresponding to old-fashioned time-ordered perturbation theory
 \cite{Baym}.
 The aim here is therefore to provide conventions and definitions along
 with a sketch of a proof, but not the details.
 A detailed derivation for a relativistic case can be found in
 Ref.~\cite{jeon} and a non-relativistic version can be found in
 Ref.~\cite{Mills}. 

 The standard Feynman rules use the momentum space Feynman
 propagator.  For our purposes, it is more convenient to use mixed
 propagators which are functions of time and spatial momentum.   
 As shown in Appendix A, the mixed propagator for both bosons and 
 fermions can be represented by 
 \be
 G_{\zeta}(\tau, {\bf k})
 =
 \int \dom{\omega} \, 
 N_\zeta(\omega)\,
 \left(
 e^{-\omega\tau}\,
 \rho^{+}_{\zeta}(k)\,\theta(\tau)\,
 +
 e^{\omega\tau}\,
 \rho^{-}_{\zeta}(k)\,\theta(-\tau)\,
 \right)
 \;.
 \ee
 Here $\omega\equiv k^0$, and the spectral densities are labeled
 $\rho^{\pm}_{\zeta}(k)$ or
 $\rho_\zeta^\pm(\omega, {\bf k})$, as convenient.
 The label $\zeta$ distinguishes boson or fermion, $\{\hbox{B, F}\}$.
 The statistical factor $N_\zeta(\omega)$ is  
 \be
 N_\zeta(\omega) = 1 + (-1)^\zeta n_\zeta(\omega)\;,
 \ee
 where $(-1)^{\rm B} \equiv 1$ and $(-1)^{\rm F} \equiv -1$
 and
 \be
 n_\zeta(\omega) = {1\over e^{\omega\beta} - (-1)^\zeta}
 \;.
 \ee
 Explicit forms for the spectral densities are not of interest in this
 section.  What is important for us is that they satisfy
 \be
 \rho^{+}_\zeta(\omega, {\bf k})
 =
 -\rho^{-}_\zeta(-\omega, {\bf k})
 \ee
 due to the periodicity or antiperiodicity of $G_\zeta(\tau, {\bf k})$ 
 in $\tau$.
 A few relevant facts about the spectral densities are given in 
 Appendices \ref{app:eucl} and \ref{app:real}. 

 The Feynman rules in this mixed space of imaginary-time and
 momentum are almost identical to the standard
 momentum space Feynman rules.  The differences are:  
 An imaginary-time $\tau_j$ labels each vertex including the external ones.  
 Each line connecting two vertices labeled by the times
 $\tau_a$ and $\tau_b$ represents a propagator
 $G_\zeta (\tau_a{-}\tau_b, |{\bf k}|)$ if the momentum flows from
 $\tau_b$ to $\tau_a$.  If the momentum flows from $\tau_a$ to $\tau_b$,
 the line represents $G_\zeta (\tau_b{-}\tau_a, |{\bf k}|)$.
 The direction of the momentum should follow the direction of charge
 flow.
 Instead of the sum over all loop frequencies, there are integrations
 over all $\tau_j$'s from $0$ to $\beta$.  
 At each vertex where an external
 field operator extracts external frequency $i\nu_j$
 an additional factor of $\exp\left( i\nu_j\tau_j \right)$ 
 is present.  
 Hence, the contribution of a connected Feynman diagram $\Gamma$
 with a total of $V{+}1$ vertices and $N{+}1$ external operator 
 insertions has the following schematic form: 
 \be
 \lefteqn{
 C_{N+1}^{(\Gamma)}(\{ {\bf q}_l, i\nu_l\})
 } & &
 \non
 & = &
 \int_0^\beta \prod_{i=0}^V d\tau_i 
 \exp\left( i\sum_{l=0}^N \nu_l\tau_l \right)\,
 \int\prod_{L\in \Gamma} {d^3 k_L\over(2\pi)^3}\,
 A_{V+1}^{N+1}
 \non
 & & {} \times
 \prod_{\alpha \in \Gamma}
 \int \dom{\omega_\alpha} \, 
 N_{\zeta_\alpha}(\omega_\alpha)\,
 \left(
 e^{-\omega_\alpha(\tau_a - \tau_b)}\,
 \rho^{+}_{\zeta_\alpha}(k_\alpha)\,\theta(\tau_a - \tau_b)\,
 +
 e^{\omega_\alpha (\tau_a - \tau_b)}\,
 \rho^{-}_{\zeta_\alpha}(k_\alpha)\,\theta(\tau_b - \tau_a)\,
 \right)
 \;.
 \label{eq:before_time_int} 
 \non
 \ee
 Here the independent spatial loop momenta are denoted by ${\bf k}_L$,
 and the quantity $A_{V+1}^{N+1}$
 includes all the factors from interaction vertices such as coupling
 constants, $\gamma$ matrices, and symmetry factors. 
 We suppress Lorentz indices, if there are any,
 since they are not relevant for the moment. 
 A given line in the Feynman diagram running between vertices acting
 at times $\tau_a$ and $\tau_b$ is denoted by 
 $\alpha$.

%
%
%
 
 We know that as the temperature goes to zero 
 the result of the time integrations must be that of old-fashioned 
 time-ordered perturbation theory.  That is,
 \be
 \lim_{\beta\to\infty}
 C_{N+1}^{(\Gamma)}(\{ {\bf q}_l, i\nu_l\})
 =
 \sum_{{\Gamma_{\sigma}} \subset \Gamma} \;
 \lim_{\beta\to\infty}
 C_{N+1}^{(\Gamma_\sigma)}(\{ {\bf q}_l, i\nu_l\})\;,
 \ee 
 where $\sigma$ represents a given permutation of the vertices and
 $\Gamma_\sigma$ represents a time-ordered diagram for that
 permutation given by 
 \be
 \lefteqn{
 \lim_{\beta\to\infty}
 C_{N+1}^{(\Gamma_\sigma)}(\{ {\bf q}_l, i\nu_l\})}
 \non
 & = &
 \int
 \prod_{L \in \Gamma} {d^3 k_L \over (2\pi)^3} \,
 \prod_{\alpha \in \Gamma}
 \left(
 \int {d\omega_{\alpha} \over 2\pi} \, 
 \theta(\omega_\alpha)\,
 \rho_{\zeta_\alpha}^{s_\sigma}(k_\alpha)
 \right)\,
 A_{V+1}^{N+1}
 \prod_{{
 \hbox{\scriptsize{\rm intervals}}
 \atop
 \hbox{$\scriptstyle{V \geq j \geq 1}$} }}
 \Bigl(
 \Lambda_j^{\sigma} -  i\nu_j^{\sigma}
 \Bigr)^{-1}
 \;.
 \label{eq:zero_temp_limit}
 \ee
 Here we omit the overall frequency-conserving Kronecker-$\delta$.
 The sign $s_\sigma = \{+, -\}$ depends on both the time
 ordering {\em and} the direction of the charge flow.  If the charge
 flows in the same direction as the time a $+$ sign is assigned,
 otherwise $s_\sigma =-$. If there are no conserved charges
 the distinction is without significance.
 The prefactor $A_{V+1}^{N+1}$ is the same as before.
 The time ordering also determines the frequency denominator
 $(\Lambda_j^{\sigma} -  i\nu_j^{\sigma})$.
 Given a time ordering $\sigma$, $\Lambda_j^\sigma$ 
 is given by the sum of frequencies of all lines crossing the
 $j$-th interval $I_j^\sigma$
 \be
 \Lambda_j^{\sigma}
 =
 \sum_{ \alpha \in I_j^\sigma }
 \omega_{\alpha}
 \;.
 \label{eq:Lambdajsigma}
 \ee
 Similarly, $\nu_j^{\sigma}$
 is the net external frequency flowing out of the diagram
 above the interval $I_j^\sigma$
 \be
 \nu_j^{\sigma}
 =
 \sum_{ V \geq l \geq j }
 \nu_l
 \;.
 \ee
 In this limit the external frequencies $\nu_l$ are continuous. 
 
 Comparing Eq.~(\ref{eq:zero_temp_limit}) and Eq.~(\ref{eq:before_time_int}) 
 one sees that one should keep all factors of 
 $N_\zeta(\omega_\alpha)$ since
 \be
 \lim_{\beta\to\infty} N_\zeta(\omega_\alpha) = \theta(\omega_\alpha)
 \;.
 \ee
 Then for Eq.~(\ref{eq:zero_temp_limit}) to be true, 
 there cannot be any additional
 factors of $\exp\left\{\pm\beta\sum_i \omega_i\right\}$
 in the finite temperature result since they will make the result
 diverge due to the fact that frequencies can be both positive and negative.
 The structure of integrand in Eq.~(\ref{eq:before_time_int}) also
 dictates that vanishing contributions such as 
 $\exp\left\{-\beta|\sum_i \omega_i| \right\}$ or $1/\beta^n$ cannot occur. 
 Then the only feasible non-zero temperature result 
 is  Eq.~(\ref{eq:zero_temp_limit}) with 
 $\theta(\omega_\alpha)$ replaced by 
 $N_{\zeta_\alpha}(\omega_\alpha)$.  
 That is, 
 \be
 {C}_{N{+}1}^{(\Gamma)} (\{ {\bf q}_l, i\nu_l\})
 =
 \sum_{{\Gamma_{\sigma}} \subset \Gamma} \;
 C_{N+1}^{(\Gamma_\sigma)}(\{ {\bf q}_l, i\nu_l\})\;,
 \ee 
 where 
 \be
 \lefteqn{C_{N+1}^{(\Gamma_\sigma)}(\{ {\bf q}_l, i\nu_l\})}
 \non
 & = &
 \int
 \prod_{L \in \Gamma} {d^3 k_L \over (2\pi)^3} \,
 \prod_{\alpha \in \Gamma}
 \left(
 \int {d\omega_{\alpha} \over 2\pi} \, 
 N_{\zeta_\alpha}(\omega_\alpha)\,
 \rho_{\zeta_\alpha}^{s_\sigma}(k_\alpha)
 \right)\,
 A_{V+1}^{N+1}
 \prod_{{
 \hbox{\scriptsize{\rm intervals}}
 \atop
 \hbox{$\scriptstyle{V \geq j \geq 1}$} }}
 \Bigl(
 \Lambda_j^{\sigma} -  i\nu_j^{\sigma}
 \Bigr)^{-1}
 \;,
 \ee
 again omitting the overall frequency-conserving Kronecker-$\delta$.

 \subsection{Retarded Correlation Functions}
 \label{sec:retarded}

 The $(N{+}1)$-point retarded function is defined by \cite{Symanzik}
 \be
 \lefteqn{ R(x; \{ x_i \}) \equiv} & &
 \non
 & &
 \sum_{\sigma}
 \theta(t{-}t_{\sigma_1})\,
 \prod_{j=1}^{N-1}
 \theta(t_{\sigma_j}{-}t_{\sigma_{j+1}})\,
 \Big\langle
 [...[
 [\hat{\varphi}(x), \hat{\varphi}_{\sigma_1}(x_{\sigma_1})], 
  \hat{\varphi}_{\sigma_2}(x_{\sigma_2})],
 ...,\hat{\varphi}_{\sigma_{N}}(x_{\sigma_{N}})] 
 \Big\rangle\,,\label{eq:retn1}
 \ee
 where 
 $t=x^0$ is the fixed largest time, $i$ runs from 1 to $N$, and 
 $\sum_{\sigma}$ indicates the sum over all possible permutation
 of $\{t_1,t_2,\ldots,t_N\}$. 
 The operator $\hat{\varphi}$ can be any field in the theory.
 When two fermionic quantities are involved, 
 the commutator becomes an anti-commutator. 
 The angular bracket represents the thermal average.
 From here on, we always denote by $v$ the vertex at fixed time $t$.

 Several authors \cite{Evans,Baier} have shown that 
 the following analytic continuation of
 the $(N{+}1)$-point imaginary-time correlation function leads 
 to the retarded function:
 \be
 i\nu_l & \to & q_l^0 + i\epsilon\ \  \hbox{for}\ \ l\ne v 
 \non
 i\nu_v & \to & q_v^0 + iN\epsilon \;,
 \ee
 where at vertices other than the fixed time vertex $v$
 the frequencies are {\em injected}, 
 and at $v$ the frequency is {\em extracted} so that
 the frequency is conserved including the small imaginary part:
 \be
 q_v^0 + iN\epsilon = \sum_{l\ne v}(q_l^0 + i\epsilon)
 \;.
 \ee
 The contribution of a Feynman diagram $\Gamma$ to the retarded
 function ${R}_{N{+}1}^{(\Gamma_v)} (\{ q_l \})$
 (subscript $v$ signifies that $v$ is the fixed vertex)
 is then:
 \begin{eqnarray}
 {R}_{N{+}1}^{(\Gamma_v)} (\{ q_l \})
 & = & \displaystyle
 \sum_{{\Gamma_{\sigma}} \subset \Gamma} \;
 \int
 \prod_{L \in \Gamma} {d^3 k_L \over (2\pi)^3} \,
 \prod_{\alpha \in \Gamma}
 \left(
 \int {d\omega_{\alpha} \over 2\pi} \, 
 N_{\zeta_\alpha}(\omega_\alpha)\,
 \rho_{\zeta_\alpha}^{s_\sigma}(k_\alpha)
 \right)
 A_{V+1}^{N+1}
 \non
 & & \qquad {} \times
 \prod_{
 {
 \hbox{\scriptsize{\rm intervals}}
 \atop
 \hbox{$\scriptstyle{V \geq j \geq v+1}$}
 }
 }
 \Bigl(
 \Lambda_j^{\sigma} -  q^{\sigma}_j + i\epsilon
 \Bigr)^{-1}
 \prod_{
 {
 \hbox{\scriptsize{\rm intervals}}
 \atop
 \hbox{$\scriptstyle{v \geq j' \geq 1}$}
 }
 }
 \Bigl(
 \Lambda_{j'}^{\sigma} -  q^{\sigma}_{j'} - i\epsilon
 \Bigr)^{-1}
 \;.
 \label{eq:R_before_sum}
 \end{eqnarray}
 The $\Lambda_j^\sigma$ were defined in Eq.~(\ref{eq:Lambdajsigma})
 and the quantities $q^{\sigma}_j$ refer to 
 the net external frequency flowing out of the diagram
 above the given interval
 \be 
 q^{\sigma}_j 
 =
 \sum_{ V \geq v_l \geq j }
 q^0_l
 \;.
 \ee
 Here we chronologically
 label the vertices by $\{t_0, \cdots, t_V\}$ 
 and the interval between $t_j$ and $t_{j-1}$ is the $j$-th interval.
 Also, a product without a factor is defined to be 1.
 Note the change in the sign of the $\epsilon$ term when crossing the $v$-th
 interval.  The sign depends on whether the interval is above or below 
 the vertex $v$ which extracts $iN\epsilon$. Intervals where 
 no external frequencies flow in or out have no $i\epsilon$ to start with, 
 but it is not hard to make the internal frequencies slightly off the real
 axis to make a consistent assignment of the imaginary
 parts \cite{jeon}. 
 
 We would like to express Eq.~(\ref{eq:R_before_sum})
 in terms of uncut and cut
 propagators in analogous fashion to the Cutkosky approach at zero 
 temperature.
 To that end, notice that if all the $+i\epsilon$'s in the
 frequency denominators were $-i\epsilon$ and 
 all the $N_{\zeta_\alpha}(\omega_\alpha)$ were $\theta(\omega_\alpha)$, 
 then the expression would be exactly that of old-fashioned real-time 
 perturbation theory.  When the time orderings are summed, 
 this expression just reproduces the Feynman diagram expression. 
 Hence when all the small imaginary parts are $-i\epsilon$,
 all we need to do to sum the time-ordered terms is to change
 the zero temperature propagator
 \be
 G^0_\zeta (k)
 = 
 \int {d\omega \over 2\pi i} \,
 \theta(\omega)\,
 \left(
 {          
  \rho_\zeta^+(\omega,|{\bf k}|)
         \over
 \omega - k^0 - i \epsilon
 }    
 +
 {          
  \rho_\zeta^-(\omega,|{\bf k}|)
         \over
 \omega + k^0 - i \epsilon
 }    
 \right) \,
 \ee
 to the finite temperature one
 \be
 G_\zeta (k)
 \equiv 
 \int {d\omega \over 2\pi i} \,
 N_\zeta(\omega)\,
 \left(
 {          
  \rho_\zeta^+(\omega,|{\bf k}|)
         \over
 \omega - k^0 - i \epsilon
 }    
 +
 {          
  \rho_\zeta^-(\omega,|{\bf k}|)
         \over
 \omega + k^0 - i \epsilon
 }    
 \right) \,
 \;,
 \ee
 see Appendix B. Then
 \be
 D_{N{+}1}^{(\Gamma)} (\{ q_l \})
 & \equiv & 
 \sum_{\Gamma_\sigma \subset \Gamma} 
 \int
 \prod_{L \in \Gamma} {d^3 k_L \over (2\pi)^3} \,
 \prod_{\alpha \in \Gamma}
 \left(
 \int {d\omega_{\alpha} \over 2\pi} \, 
 N_{\zeta_\alpha}(\omega_\alpha)\,
 \rho_{\zeta_\alpha}^{s_\sigma}(\omega_\alpha,|{\bf k}_\alpha|)
 \right)
 A_{V+1}^{N+1}
 \non
 & & \qquad\qquad {} \times
 \prod_{
 {
 \hbox{\scriptsize{\rm intervals}}
 \atop
 \hbox{$\scriptstyle{V \geq j \geq 1}$}
 }
 }
 \Bigl(
 \Lambda_j^{\sigma} -  q^{\sigma}_j - i\epsilon
 \Bigr)^{-1}
 \non
 & = &
 -i(i)^{V+1}
 \int
 \prod_{L \in \Gamma}
 {d^4 k_{L} \over (2\pi)^4} \,
 A_{V+1}^{N+1}
 \prod_{\alpha \in \Gamma}
 G_{\zeta_\alpha} (k_{\alpha}) 
 \;.
 \label{eq:no_cut}
 \ee
 The factor $-i(i)^{V+1}$ comes from having $V$ integrals with
 $-i\epsilon$.  The phase $i^{V+1}$ can be absorbed into
 $A_{V+1}^{N+1}$ by adding a factor of $i$ to the coupling constant
 at each vertex as in the usual diagram rules, {\it e.g.} see
 Peskin and Schroeder \cite{peskin} whose conventions we follow.

 We would like to apply this argument to each of the $\pm i\epsilon$
 parts of the product in ${R}_{N{+}1}^{(\Gamma_v)} (\{ q_l \})$.
 In order to do so, the two parts must be separated from each other so
 that the lines common to both parts can be regarded as external ones. 
 This is achieved by using the identity
 \begin{equation}
 0
 =
 \prod_{j=v+1}^{V} \Bigl(U_j {-} i\epsilon \Bigr)^{-1}
 -
 \prod_{j=v+1}^{V} \Bigl(U_j {+} i\epsilon \Bigr)^{-1}
 -i
 \sum_{l=v+1}^{V} 2\pi\delta (U_l)
 \prod_{k=l+1}^{V}  \Bigl(U_k {+} i\epsilon \Bigr)^{-1}
 \prod_{j=v+1}^{l-1} \Bigl(U_j {-} i\epsilon \Bigr)^{-1}
 \label{eq:terms_in_disc}
 \;.
 \end{equation}
 Here all $U_j$'s are real, a sum without a summand is 0 and a product
 without a factor is~1.  This identity follows directly from 
 $2\pi i\delta(x) = (x-i\epsilon)^{-1} - (x+i\epsilon)^{-1}$.
 We refer to the lines corresponding to the interval in $\delta(U_l)$ as
 the cut lines.  
 
 Using the above identity and performing the relevant resummation of time
 ordered terms on both sides of the cut, we obtain 
 \begin{equation}
 {R}_{N{+}1}^{(\Gamma_v)} (\{ q_l \})
 =
   {D}_{N{+}1}^{(\Gamma)}(\{ q_l \})
    +
   \sum_{{\Gamma\llap{${\scriptstyle -}$}}_v \subset \Gamma}
   {R}_{N{+}1}^{({\Gamma\llap{${\scriptstyle -}$}}_v)} (\{ q_l \})
 \;,
 \label{eq:main_result}
 \end{equation}
 where
 \be
 R^{({\Gamma\llap{${\scriptstyle -}$}}_v)}_{N+1} (\{ q_l \})
 & = &
 -i (-i)^{V_+}(i)^{V_-}
 \int
 \prod_{L \in \Gamma}
 {d^4 k_{L} \over (2\pi)^4} \,
 A_{V+1}^{N+1}
 \prod_{{
   \hbox{\scriptsize{\rm cut lines}}
   \atop
   \hbox{\scriptsize{\it c}}
  }}
 \left(
 N_{\zeta_c}(k^0_{c}) \,
 \rho_{\zeta_c}^{s_c}( k_{c} )
 \right)
 \non
 & & \qquad\qquad\qquad\qquad {} \times
 \prod_{\alpha \in \Gamma_{+}}
 G_{\zeta_\alpha} (k_{\alpha})^* \,
 \prod_{\alpha' \in \Gamma_{-}}
 G_{\zeta_{\alpha'}} (k_{\alpha'}) 
 \;,
 \label{eq:contribution.of.cut}
 \ee
 and the sum is over the cut diagrams ${\Gamma\llap{$-$}}_v$.  
 The uncut propagator $G_\zeta$ corresponds to the lines 
 in the side $\Gamma_-$ which contains $v$, 
 and $G_\zeta^*$ corresponds to the lines in the other
 side $\Gamma_+$.
 For the cut lines the direction of time flow is {\em from} 
 $\Gamma_-$ {\em to} $\Gamma_+$.
 Here $V_\pm$ denotes number of vertices in $\Gamma_\pm$.
 Again, the factors $(-i)^{V_+}(i)^{V_-}$
 can be absorbed into $A_{V+1}^{N+1}$ by including 
 a factor of $i$ in the contribution of each vertex in $\Gamma_-$ 
 and a factor of $-i$ in the contribution of each vertex in $\Gamma_+$. 
 
 So far, nothing has been dependent on the presence of chemical
 potentials.  When chemical potentials are present, 
 Eq.~(\ref{eq:main_result}) corresponds to the retarded function of
 fields $\hat{\varphi}_K(t) = e^{i\hat{K}t}\hat{\varphi}e^{-i\hat{K}t}$ 
 with $\hat{K} = \hat{H} - \sum_a \mu_a \hat{Q}_a$.
 As explained in Sec.~\ref{sec:intro}, there is a simple
 relation between $\hat{\varphi}_K$ and $\hat{\varphi}_H$,
 $\hat{\varphi}_K(t) = e^{i\mu_\varphi t}\hat{\varphi}_H(t)$.
 Hence, to convert a retarded function of the $\hat{\varphi}_K$ to the
 corresponding retarded function of the $\hat{\varphi}_H$, all one has 
 to do is to shift each external frequency from $q$ to $q - \mu_\varphi$ 
 if it corresponds to the particle of the species
 and to $q + \mu_\varphi$ if it corresponds to the anti-particle. 
 These shifts have important simplifying consequences.  
 Due to charge conservation, not only 
 energy-momentum, but also chemical potential has to be conserved at
 each vertex. 
 Hence, shifting the external frequency $q^0 \to q^0 - \mu_\varphi$ 
 is equivalent to shifting
 {\em all} internal frequencies 
 $l^0 \to l^0 - \mu_\alpha$ according to the species $\alpha$.
 
 The significance 
 of these internal frequency shifts is that they remove the
 chemical potential dependence from the spectral densities.  
 For instance, when the chemical potential is non-zero, the spectral
 density of the bosonic propagator appearing in Eq.~(\ref{eq:R_before_sum})
 is (see Appendix~\ref{app:eucl})
 \be
 \rho^+_{\rm B}(k^0)
 =
 {\rm sign}(k^0+ \mu)\,
 2\pi\,\delta\left( (k^0+ \mu)^2 - E_k^2 \right)
 \;,
 \ee
 where $E_k= \sqrt{{\bf k}^2 + m^2}$.
 By shifting $k^0 \to k^0 - \mu$, 
 $\rho^+_{\rm B}(k^0-\mu)$
 becomes independent of $\mu$, while the same change will make the
 statistical factor become $n_{\rm B}(\omega - \mu)$.\footnote{
	 These internal shifts also have a simplifying effect on
	 derivative couplings.  The derivative 
	 $\partial_t$ present at an interaction vertex yields
	 $l^0 + \mu$ which becomes simply $l^0$ after
	 the shift of internal frequencies.
}

 For small chemical potentials such that $m_\zeta > |\mu_\zeta|$,
 even more simplification is possible because in that case,
 $\theta(E \pm \mu) = \theta(E)$ and $\theta(-E \pm \mu) = \theta(-E)$.
 Then 
 \be
 N_\zeta(k^0 - \mu)
 = 
 \theta(k^0) + (-1)^\zeta {\rm sign}(k^0) n_\zeta(|k^0 - \mu|)
 \;,
 \ee 
 when $k^0 = \pm E_k$.
 Consequently,
 the chemical potential only appears in the statistical factors.  
 In this way we achieve the separation of the statistical part and the
 purely dynamic part of the particle propagation.  (For a slightly different
 approach, see \cite{Dashen}).  

 We can now state diagrammatic rules for calculating an arbitrary
 $N$-point retarded function with small chemical potentials 
 $(|\mu_\zeta| < m_\zeta)$, see Appendix~\ref{app:real} for further 
 details. For convenience, we denote the thermal phase 
 space factor for scalar bosons by
 \be
 \Gamma_{\rm B}(k)
 =
 n_B(|k^0 -\mu|) \,2\pi\,\delta(k^2 - m^2)
 \label{eq:gamma_zeta_B}
 \ee
 and for spin-$\frac{1}{2}$ fermions
 \be
 \Gamma_{\rm F}(k)
 =
 -n_{\rm F}(|k^0 -\mu|)\,
 (k^\mu\gamma_\mu + m)
 \,2\pi\,\delta(k^2 - m^2)
 \label{eq:gamma_zeta_F}
 \;.
 \ee
 Define the cut propagators for scalar particles 
 \be
 \Delta_{\rm B}^{\pm}(k) 
 =
 \theta(\pm k^0) \, 2\pi\,\delta(k^2 - m^2)
 + \Gamma_{\rm B}(k)
 \;,
 \label{eq:Del_B}
 \ee
 and for fermions, 
 \be
 \Delta_{\rm F}^{\pm}(k) 
 =
 \theta(\pm k^0) \, 2\pi\, (k^\mu\gamma_\mu + m)\, \delta(k^2 - m^2)
 + \Gamma_{\rm F}(k)
 \;.
 \label{eq:Del_F}
 \ee
 For scalar particles, the uncut propagators are given by 
 \be
 G_{\rm B}(k)
 =
 {i \over k^2 - m^2 + i\epsilon}
 +
 \Gamma_{\rm B}(k)
 \ \ \hbox{and}\ \ 
 G^*_{\rm B}(k)
 = 
 {-i \over k^2 - m^2 - i\epsilon}
 +
 \Gamma_{\rm B}(k)
 \ee
 and for spin-$\frac{1}{2}$ fermions, they are given by
 \be
 G_{\rm F}(k)
 = 
 {i (k^\mu \gamma_\mu + m) \over k^2 - m^2 + i\epsilon}
 +
 \Gamma_{\rm F}(k)
 \ \ 
 \hbox{and}
 \ \  
 G^*_{\rm F}(k)
 = 
 {-i (k^\mu \gamma_\mu + m) \over k^2 - m^2 - i\epsilon}
 +
 \Gamma_{\rm F}(k)
 \label{eq:Fermion_G_star}
 \;.
 \ee
 For gauge bosons, set $m = 0$ in each of the scalar particle propagator,
 and multiply by $(-g_{\mu\nu})$ (Feynman gauge).
 Note that {\em all} of the cut and uncut propagators can be written as
 a zero-temperature, zero-density part plus a common thermal phase space
 factor. 
 Not surprisingly,
 the zero temperature parts of the propagators all coincide
 with the zero-temperature Cutkosky rule propagators \cite{t'Hooft}.

 We call the region where the fixed vertex $v$ sits the unshaded region
 and the other half the shaded region.  
 The rules to calculate the retarded functions of $\hat\varphi_H$ are:
 \begin{enumerate}
 \item
 Draw all topologically distinct cut diagrams including totally uncut
 ones, keeping the largest time always on the unshaded side. 
 Disconnected pieces produced by the cutting are allowed.  

 \item 
 Assign momenta to the lines according to the flow of the conserved 
 charges, if there are any. 
 
 \item
 Use the usual Feynman rules for the unshaded side
 assigning $G_\zeta(k)$ to the uncut lines.  

 \item
 Use the complex conjugate Feynman rules for the shaded side assigning
 $G_\zeta^*(k)$ to the uncut lines.
 The $\gamma$ matrices are not to be complex conjugated.
 
 \item
 If the momentum $k$ of a cut line crosses from the unshaded region 
 to the shaded region, assign $\Delta_\zeta^+(k)$.
 
 \item
 If the momentum $k$ of a cut line crosses from the shaded region 
 to the unshaded region, assign $\Delta_\zeta^-(k)$.
 
 \item 
 Divide by the symmetry factor if applicable.
 There is an overall factor of $-i$.  

 \end{enumerate}
 Equation (\ref{eq:main_result}) and the above diagrammatic rules
 are the main results of this section.  
 As shown in Appendix~\ref{app:derivative},
 this same set of rules also apply when there are derivative couplings.

\begin{figure}[t]
 \setlength{\unitlength}{1cm}
 \begin{center}
 \leavevmode
 \epsfxsize=9cm
 \epsfbox{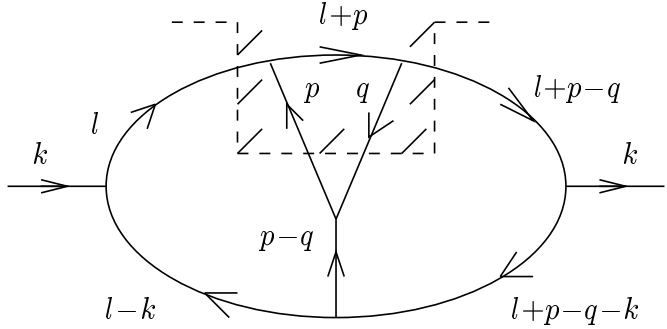}
 \end{center}
 \caption{A typical finite temperature cut diagram in a scalar
 $g\phi^3$ theory. Each cut line
 contributes $\Delta_{\rm B}^+$ or $\Delta_{\rm B}^-$ 
 according to its orientation.
 In the unshaded (shaded) region, an uncut propagator 
 contributes $G_{\rm B}$ ($G_{\rm B}^*$),
 and an interaction vertex contributes a factor of $-ig$ $(ig)$.
 }
 \label{fig:example}
 \end{figure}

 As an example of applying the rules,
 consider a scalar theory with a $-g\phi^3$ interaction in the Lagrangian.
 The diagram shown in Fig.~\ref{fig:example}
 yields the expression
 \be
 C^\scrpt{fig.\protect{\ref{fig:example}}}(k) 
 & = &
 (-i)
 (-ig)^4(ig)^2
 \int {d^4 l \over (2\pi)^4}\, {d^4 p\over (2\pi)^4}\,
 {d^4 q \over (2\pi)^4}\,
 \Delta_{\rm B}^+(l)\,
 \Delta_{\rm B}^+(p)\,
 \Delta_{\rm B}^-(q)\,
 \Delta_{\rm B}^-(l{+}p{-}q)\,
 \non
 & & \qquad\qquad\qquad\qquad {} \times
 G_{\rm B}(l{+}p{-}q{-}k)\,
 G_{\rm B}(l{-}k)\,
 G_{\rm B}(p{-}q)\,
 G^*_{\rm B}(l{+}p)
 \;.
 \ee
 Note that at $T=0$, the contribution of
 this diagram is zero because the cut part
 then represents a process where four physical particles annihilate 
 each other into the vacuum.
 At non-zero
 temperature, this diagram gives a non-zero contribution because it 
 can also represent scattering between physical particles in the 
 thermal medium.

 \section{ Multiple Scattering Expansion of the Self-Energy }
 \label{sec:self_energy}

 In this section, we present the
 multiple scattering expansion for the thermal contribution to the self-energy.
 The retarded self-energy is, of course, a retarded 2-point function.
 From the previous section (see also Appendix~\ref{app:real})
 we know that as long as $m_a > |\mu_a|$, all propagators satisfy
 \be
 D_\zeta(k) = D_\zeta^0(k) +  \Gamma_\zeta(k)
 \label{eq:Dzeta}
 \ee
 where $\Gamma_\zeta(k)$ 
 is the thermal phase space factor common to all four cut and uncut
 propagators,
 and $D_\zeta^0(k)$ is the zero temperature, zero density propagator.  
 
 Using Eq.~(\ref{eq:Dzeta})
 we can now expand the expression (\ref{eq:main_result}) for the
 self-energy in the number of $\Gamma_\zeta$'s.
 The coefficients in this expansion 
 involve {\em only the zero temperature propagators}.  
 Hence, the coefficient function is either a
 zero-temperature Feynman diagram 
 or a zero-temperature Cutkosky diagram. 
 External momenta for this coefficient function are provided 
 by the self-energy momentum, $k$, and the thermal particle momenta 
 from the $\Gamma_\zeta$'s.  

 We would like to relate the coefficients of the expansion to the
 physical zero-temperature scattering amplitudes.  
 In order to do so, we need to ensure that in the expansion we consider:
 \begin{enumerate}
 \item
 Symmetry factors are all correctly accounted for.
 \item
 Disconnected parts cancel. 
 \item 
 Self-energy insertions do not cause divergences. 
 \item
 Additional polarization factors needed 
 for the external thermal particles are all correctly provided. 
 \item
 When fermions are involved, the overall sign for an individual Feynman
 diagram is correctly produced.
 \end{enumerate}
 We present the result of the expansion first and deal with the above
 points later in Subsecs.~\ref{subsec:sym}--\ref{subsec:insertions}.
 
 We emphasize that
 the formal expansion in the number of $\Gamma_\zeta$'s
 can be always made.  
 To be useful, however, the series must be truncated.
 When the temperature and the density are low so that 
 $T \ll |m_a \pm \mu_a|$ for all particle species $a$,
 one may order the expansion with respect to the relative strengths of
 $e^{-(m_a -\mu_a)/T}$ and $e^{-(m_a +\mu_a)/T}$.  The first few terms of
 such a virial expansion will be a good approximation as long as
 the densities and the temperature stay low.
 As the densities and the temperature grow so that the minimum of 
 $|m_a \pm \mu_a|$ is no longer small compared to the temperature, we
 lose control of the approximation because, potentially,
 an infinite number of terms in the expansion become important.
 
\begin{figure}[t]
 \setlength{\unitlength}{1cm}
 \begin{center}
 \leavevmode
 \epsfxsize=14cm
 \epsfbox{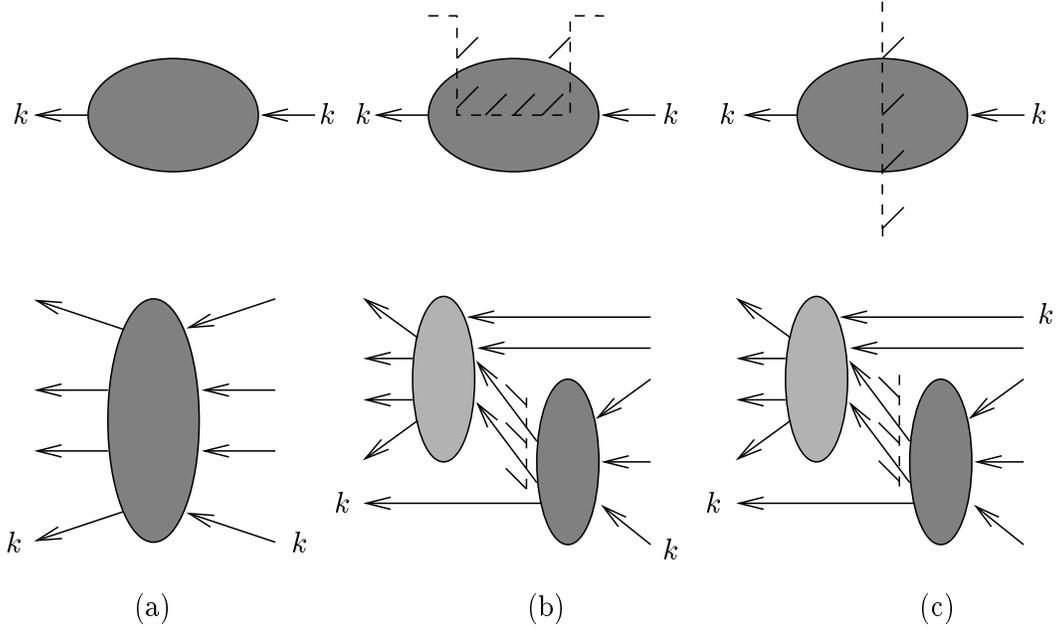}
 \end{center}
 \caption{Diagrams that contribute to the
 self-energy. The upper row is illustrated in more detail in the lower
 row.}
 \label{fig:diagrams}
 \end{figure}

 The diagrams that provide the thermal contributions to the retarded 
 self-energy, $\Sigma^\scrpt{Ret}_T$, can be 
 separated into three groups, as depicted in Fig. \ref{fig:diagrams}.
 Group (a) contains no cut,
 group (b) contains 
 cuts that do not separate the two
 external vertices and group (c) contains cuts that do separate
 the two external vertices.
 
 Consider first the diagrams depicted in
 Fig.~\ref{fig:diagrams}(a).
 The diagrams in this group are solely made of the uncut propagator 
 $G_\zeta(l) = G_\zeta^0(l) + \Gamma_\zeta(l)$, where $G_\zeta^0(l)$ is
 the zero temperature Feynman propagator and $\Gamma_\zeta(l)$ is the
 thermal phase space factor.
 When the expansion in the number of $\Gamma_\zeta$ is made, 
 the coefficients are obviously the zero temperature Feynman diagrams
 whose external momenta are supplied
 by the self-energy momentum and the thermal particle momenta
 from the $\Gamma_\zeta$'s.  
 That is, the coefficients correspond 
 to matrix elements of the scattering operator ${\cal T}$, with the 
 $S$-matrix given by ${\cal S} = 1 + i{\cal T}$.
 Hence, the expansion of the first group of the diagrams can be written
 as
 \be
 \Sigma^\scrpt{(a)}_T(k)
 & = &
 -
 \sum_{n \ge 2, \sigma}
 {1\over S_{\{l_i^\sigma \}}}
 \int \prod_{i=1}^{n-1}d\Gamma_i^\sigma\,
 \langle k, \{l_i^\sigma \} |  
 {\cal T} 
 | k, \{l_i^\sigma \} \rangle_\scrpt{conn}
 \non
 & & {}
 + \left( \hbox{disconnected diagrams} \right)
 \;,\label{eq:sig1}
 \ee
 where 
 \be
 d\Gamma_i^\sigma\equiv\frac{d^3l_i}{(2\pi)^32E_i}
 n_{\zeta_i}(E_i-\sigma\mu_i)\;,
 \ee
 with $E_i=\sqrt{{\bf l}_i^2+m_i^2}$ and $\sigma=+1$ for a particle and $-1$
 for an antiparticle. In Eq. (\ref{eq:sig1}) the symmetry factor is denoted 
 by $S_{\{l_i^\sigma \}}$; this will be discussed in the 
 following subsection.

 Next consider the diagrams of  Fig.~\ref{fig:diagrams}(b) shown
 schematically in the upper row and in more detail in the lower row.
 To express these diagrams 
 in terms of the scattering operator,
 we define an operator ${\cal T}_k$ such that
 \be
 \langle k, \{ l_i^\sigma \} |
  {\cal T}_k 
 | \{ p_i \} \rangle
 & \equiv &
 \langle k, \{ l_i^\sigma \} |
 {\cal T} 
 | \{ p_i \} \rangle
 -
 (\hbox{disconnected parts involving $\delta(k-p_i)$})
 \non
 \inline{and}
 \langle \{ p_i \} |
  {\cal T}_k 
 | k, \{ l_i^\sigma \} \rangle
 & \equiv &
 \langle \{ p_i \} |
 {\cal T} 
 | k, \{ l_i^\sigma \} \rangle
 -
 (\hbox{disconnected parts involving $\delta(k-p_i)$})
 \;.
 \label{eq:def_Tk}
 \ee
 We also define
 \be
 {\Delta}_k \equiv {\cal T} - {\cal T}_k
 \;. \label{eq:Deltadef}
 \ee
 The coefficients originating 
 from expanding a cut diagram in this second group
 must also contain a cut with
 the same restriction that the two external vertices should always be
 in the unshaded region.
 Hence, the coefficient functions are the zero temperature Cutkosky
 diagrams with the same restriction.
 We know that the zero temperature Cutkosky diagrams represent matrix
 element of ${\cal T}^\dagger {\cal T}$. 
 Hence, the expansion of this group of diagrams must be 
 \be
 \Sigma^\scrpt{(b)}_T(k)
 & = &
 i 
 \sum_{n \ge 2, \sigma}
 {1\over S_{\{l_i^\sigma \}}}
 \int \prod_{i=1}^{n-1}d\Gamma_i^\sigma\,
 \sum_{\{p_j\}}
 \langle \{l_i^\sigma \} |  
  {\cal T}^\dagger 
 | \{ p_j \} \rangle
 \langle k, \{ p_j \} |
  {\cal T} 
 | k, \{l_i^\sigma \} \rangle|_\scrpt{conn}
 \non
 & & {}
 + \left( \hbox{disconnected diagrams} \right)
 \non
 & = &
 i 
 \sum_{n \ge 2, \sigma}
 {1\over S_{\{l_i^\sigma \}}}
 \int \prod_{i=1}^{n-1}d\Gamma_i^\sigma\,
 \langle k, \{l_i^\sigma \} |  
  {\Delta}_k^\dagger 
  {\cal T}_k 
 | k, \{l_i^\sigma \} \rangle|_\scrpt{conn}
 \non
 & & {}
 + \left( \hbox{disconnected diagrams} \right)
 \;.
 \ee
 
 The third group of diagrams
 contain a cut that separates the external vertices as shown in
 Fig.~\ref{fig:diagrams}(c).
 Noting that the external frequency $k$ flows out of the unshaded region
 and the cut-line frequencies flow into the shaded region, 
 we can write 
 \be
 \Sigma^\scrpt{(c)}_T(k)
 & = &
 i 
 \sum_{n \ge 2, \sigma}
 {1\over S_{\{l_i^\sigma \}}}
 \int \prod_{i=1}^{n-1}d\Gamma_i^\sigma\,
 \sum_{\{p_j\}}
 \langle \{l_i^\sigma \} |  
  {\cal T}^\dagger 
 | k, \{ p_j \} \rangle
 \langle k, \{ p_j \} |
  {\cal T} 
 | \{l_i^\sigma \} \rangle|_\scrpt{conn}
 \non
 & & {}
 + \left( \hbox{disconnected diagrams} \right)
 \non
 & = &
 i 
 \sum_{n \ge 2, \sigma}
 {1\over S_{\{l_i^\sigma \}}}
 \int \prod_{i=1}^{n-1}d\Gamma_i^\sigma\,
 \langle k, \{l_i^\sigma \} |  
  \Delta_k^\dagger \Delta_k 
 | k, \{l_i^\sigma \} \rangle|_\scrpt{conn}
 \non
 & & {}
 + \left( \hbox{disconnected diagrams} \right)
 \;.
 \ee
 
 Ignoring the disconnected diagrams which
 will later be shown to cancel, the three contributions yield
 \be
 \Sigma^\scrpt{Ret}_T(k)
 & = &
  \Sigma^\scrpt{(a)}_T(k)
  +
  \Sigma^\scrpt{(b)}_T(k)
  +
  \Sigma^\scrpt{(c)}_T(k)
 \non
 & = &
 -
 \sum_{n \ge 2, \sigma}
 {1\over S_{\{l_i^\sigma \}}}
 \int \prod_{i=1}^{n-1}d\Gamma_i^\sigma\,
 \langle k, \{l_i^\sigma \} |  
 \left(
  {\cal T} 
  -i{\Delta}_k^\dagger {\cal T}
 \right)
 | k, \{l_i^\sigma \} \rangle_\scrpt{conn}
 \non
 & = &
 -
 \sum_{n \ge 2, \sigma}
 {1\over S_{\{l_i^\sigma \}}}
 \int \prod_{i=1}^{n-1}d\Gamma_i^\sigma\,
 \langle k, \{l_i^\sigma \} |  
  {\cal T}_k^\dagger {\cal S} 
 | k, \{l_i^\sigma \} \rangle_\scrpt{conn}
 \;,
 \label{eq:self_energy}
 \ee
 where we used 
 $
 \langle k, \{l_i^\sigma \} |  
   {\cal T}_k 
 | k, \{l_i^\sigma \} \rangle_\scrpt{conn}
 =
 \langle k, \{l_i^\sigma \} |  
   {\cal T} 
 | k, \{l_i^\sigma \} \rangle_\scrpt{conn}
 $,
 Eq.~(\ref{eq:Deltadef}) and the identity
 ${\cal T} = {\cal T}^\dagger {\cal S} = {\cal T}^\dagger (1 + i{\cal T})$. 

 We can now show that Eq.~(\ref{eq:shuryak}) is the lowest order
 approximation of the expansion (\ref{eq:self_energy}). 
 Without self-energy insertions, the matrix element
 $ 
 \langle k, l |  
  {\Delta}_k^\dagger {\cal T}
 | k, l \rangle_\scrpt{conn}
 $
 vanishes {\em provided} that the momentum $l$ represents 
 a stable particle which cannot decay at $T = 0$.
 The lightest particle in a theory is certainly stable.  Hence, 
 Eq.~(\ref{eq:shuryak}) holds at the lowest order in the density
 expansion.
 
 To check the consistency of
 Eq.~(\ref{eq:self_energy}), consider the imaginary part,
 \be
 {\rm Im}\,\Sigma^\scrpt{Ret}_T(k)
 & = &
 -
 {1\over 2} 
 \sum_{n \ge 2, \sigma}
 {1\over S_{\{l_i^\sigma \}}}
 \int \prod_{i=1}^{n-1}d\Gamma_i^\sigma\,
 \langle k, \{l_i^\sigma \} |  
 \left(
  {\cal T}^\dagger_k {\cal T}_k 
  -
  {\Delta}_k^\dagger \Delta_k 
 \right)
 | k, \{l_i^\sigma \} \rangle_\scrpt{conn}
 \;.
 \label{eq:im_self}
 \ee
 Here we used the unitarity condition
 ${\rm Im}\,{\cal T} = {1\over 2}{\cal T}^\dagger {\cal T}$ to get 
 Eq.~(\ref{eq:im_self}) from the second line of
 Eq.~(\ref{eq:self_energy}). 
 The cut diagrams leading to expression (\ref{eq:im_self}) must
 contain cuts that separate the two external vertices.   
 Also the diagrams corresponding to ${\cal T}_k^\dagger {\cal T}_k$
 must have the external momentum $k$
 entering the unshaded region, while those
 corresponding to $\Delta_k^\dagger \Delta_k$ must have the
 external momentum $k$ exiting the unshaded region.  
 In Ref.~\cite{jeon} it is shown that
 these cut diagrams are exactly the diagrams that gives the spectral
 density of the 2-point correlation function.
 The retarded self-energy is the negative of
 a 2-point retarded correlation function with a real
 spectral density (see Appendix~\ref{app:eucl}).  
 This implies that 
 \be
 {\rm Im}\,\Sigma^\scrpt{Ret}(k)
 =
 -{\textstyle{1\over 2}}\, \chi^\dum_\Sigma(k)
 \;,
 \ee
 where $\chi^\dum_\Sigma(\omega)$ is the spectral density for the
 self-energy.  This is consistent with Eq.~(\ref{eq:im_self}). 

 Eq. (\ref{eq:self_energy}) can quickly be used to make contact with the thermal 
 ($T>0$) part of the thermodynamic grand potential; for a more rigorous 
 discussion see Appendix~\ref{app:omega}. We need to close off the diagrams
 of Fig.~\ref{fig:diagrams} by including a propagator for the external 
 line $k$. The zero temperature part will give a contribution that can be 
 included in the scattering matrices, while the finite temperature part
 gives an additional $d\Gamma$ integration and the symmetry factor must be 
 appropriately adjusted. We note that in Eq. (\ref{eq:self_energy})
 ${\cal T}_k^\dagger {\cal S}
 ={\cal T}_k^\dagger \sum_{m=0}^\infty\left(i{\cal T}^\dagger\right)^m$.
 When the propagator for $k$ is included, ${\cal T}_k^\dagger$ is no
 longer distinguished from ${\cal T}^\dagger$, but we must include a factor 
 of $(m+1)^{-1}$ in the sum in order to avoid multiple counting. Thus we have 
 $-i\sum_{m=0}^\infty\left(i{\cal T}^\dagger\right)^{m+1}/(m+1)
 =i\ln\left(1-i{\cal T}^\dagger\right)$. Thus the thermal part of the grand 
 potential can be written
 \be
 {\Omega_T\over V}&=& -i
 \sum_{n \ge 2, \sigma}
 {1\over S_{\{l_i^\sigma \}}}
 \int \prod_{i=1}^{n}d\Gamma_i^\sigma\,
 \langle \{l_i^\sigma \} |  
 \ln(1 - i{\cal T}^{\dagger})
 | \{l_i^\sigma \} \rangle_\scrpt{conn}
 \non
 &=& i \sum_{n \ge 2, \sigma}
 {1\over S_{\{l_i^\sigma \}}}
 \int \prod_{i=1}^{n}d\Gamma_i^\sigma\,
 \langle \{l_i^\sigma \} |  
 \ln(1 + i{\cal T})
 | \{l_i^\sigma \} \rangle_\scrpt{conn}
 \;, \label{eq:omnor}
 \ee
 where $V$ denotes the volume of the system and in the last step we have 
 used the reality of $\Omega$. We have also 
 assumed that the physical mass of the particles is used in the propagators 
 so that diagrams with a single thermal weighting are superfluous.
 Equation (\ref{eq:omnor}) agrees with Norton's version \cite{Norton} of the 
 expression originally given by Dashen, Ma and Bernstein \cite {Dashen}.


 \subsection{Symmetry factors}
 \label{subsec:sym}

\begin{figure}[t]
 \begin{center}
 \leavevmode
 \epsfxsize=7cm
 \epsfbox{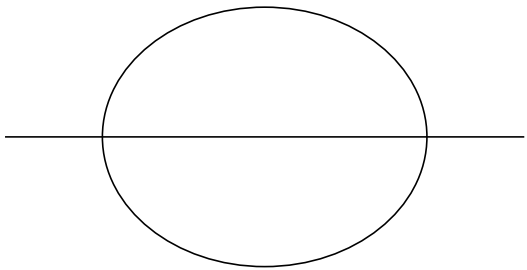}
 \end{center}
 \caption{A 2-loop diagram in scalar $\lambda\phi^4$ theory.}
 \label{fig:2_loop_mother}
 \end{figure}

 There are two issues involving the symmetry factor, $S_{\{l_i^\sigma \}}$,
 arising in the equations above.  One involves the
 symmetry factor due to the self-interaction.  For instance, the two
 loop diagram in a $\lambda\phi^4$ theory
 shown in Fig.~\ref{fig:2_loop_mother} has the symmetry 
 factor of $S_\scrpt{2-loop} = 3!$ since
 the three internal lines are equivalent to each other.  
 Our density expansion
 inevitably reduces the symmetry of this diagram.  
 One needs to show that the expansion in the number of $\Gamma_\zeta$
 automatically generates 
 the symmetry factor associated with the reduced symmetry in such cases.
 The second issue involves identical particles.  
 When there are $m$ identical particles in the final state,
 the $m$-body phase-space includes a factor of $1/m!$.
 Our expansion must also automatically account for this factor. 
 
 To see that the symmetry factors are correctly generated,
 we label each line in a diagram
 by $l_\alpha$ and each vertex by $\tau_a$ and regard
 them as distinguishable.  
 The symmetry group $G$ of a given Feynman diagram then
 represents the permutations of the internal 
 lines and the vertices that does not
 change the shape of the diagram \cite{Itzykson}.  
 The order $q_G$ of this group is the symmetry factor for the diagram. 
 For example, the symmetry group of
 the two-loop diagram in Fig.~\ref{fig:2_loop_mother} 
 is the full permutation group of the three identical lines. 
 Hence, $S_\scrpt{2-loop} = q_G = 3!$.
 
 Consider a cut diagram for the self-energy.
 The density expansion essentially amounts 
 to the sum of all possible ways of mixing the zero-temperature propagators 
 and the thermal phase space factors without totally disconnecting
 the diagram. In such ``divided diagrams"
 the lines and the vertices in the original 
 diagram are apportioned into the thermal part 
 (which solely consists of $\Gamma_\zeta$'s)
 and the zero-temperature part.
 
 In general this partitioning 
 will reduce the symmetry group of the original group $G$ to a subgroup $S$. 
 The reduced symmetry group $S$ is then a product group consisting of
 the symmetry groups $S_{\rm z}$ 
 and $S_{\rm t}$ of the two parts.
 Here the subscript `z' represents the zero-temperature part, and the
 subscript `t' represents the thermal part. 
 In other words the symmetry group is now,
 \be
 S = S_{\rm z} \otimes S_{\rm t}
 \;.
 \ee
 The order of this product group is of course 
 $q_S = q_{\rm z} \times q_{\rm t}$,
 where $q_{\rm z}$ and $q_{\rm t}$ are the order of the subgroups
 $S_{\rm z}$ and $S_{\rm t}$, respectively.
 We must show that the density expansion automatically generates
 the symmetry factor 
 $q_S = q_{\rm z}\times q_{\rm t}$
 associated with the reduced diagram.

 It is well known in group theory that {\em the order of a subgroup
 is a factor of the order of the full group} \cite{mathews}.  
 That is, $q_G/q_S = r$ is an integer.  
 We would like to show that there are
 exactly $r$ different partitions of the original lines and vertices that
 result in the same divided diagram.  
 Then we show that the correct symmetry factor
 is produced for each individual diagram in our expansion. 

 To see how it works out,
 suppose the symmetry group $G$ of the original diagram 
 is of order $q_G$, and when divided into zero-temperature 
 and thermal parts, 
 the symmetry group reduces to a subgroup $S$ of order $q_S$. 
 Consider the right coset,
 \be
 G/S = \{ S g_1, S g_2, \cdots, S g_{q_G} \}\;,
 \ee
 where each $g_i \in G$ is a permutation acting on the internal vertices 
 and the lines that does not change the shape of the original diagram.
 Since the identity must be an element of $G$
 and the $S g_i$'s are either identical or disjoint,
 we can rewrite
 \be
 G/S = \{ S_1, S_2, \cdots, S_r \}\;,
 \ee
 where $S_1 = S$, $S_i$'s are all disjoint, each $S_i$ has $q_S$ elements,
 and $r = q_G/q_S$ \cite{mathews}.
 
 Operating with $s_i \in S_1$ on the diagram does not change 
 the partition of the lines and vertices 
 since $S_1 = S$ is the symmetry group of the divided diagram. 
 However, the other $S_i$'s
 must contain at least one element that corresponds to a different
 partition.
 Then there exist $h_i$'s such that $h_i\in G$ and
 \be
 S_i = S h_i = \{ h_i, s_2 h_i, s_3 h_i, \cdots, s_{q_S} h_i \}
 \;.
 \ee
 Here, $s_i \in S$, $s_1 = I$ and $h_1=I$.
 Since $S_i$ and $S_j$ are disjoint for $i\ne j$, we have $h_i \ne h_j$.
 The number of $h_i$'s is then equal to $r = q_G/q_S$. 
 Also each element of $S_i$ is related to $h_i$ 
 by an element of $S$.  This implies that all elements in $S_i$
 correspond to the same partition of the lines and vertices.
 Hence the total number of 
 inequivalent partitions of the internal lines and vertices  
 is exactly the same as the number of $h_i$'s.
 Consequently, when our density expansion is made, the coefficient
 of a divided diagram is $r/q_G = 1/(q_{\rm z}\times q_{\rm t})$.
 Thus the symmetry factor associated with
 the symmetry of the divided diagram is always correctly produced by the
 density expansion.
 
\begin{figure}[t]
 \begin{center}
 \leavevmode
 \epsfxsize=10cm
 \epsfbox{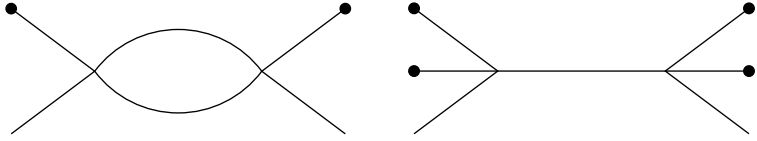}
 \end{center}
 \caption{Density expansion of the 2-loop diagram in
 Fig.~\protect\ref{fig:2_loop_mother}.  The dot at the end of a line
 denotes that the line corresponds to a thermal particle.}
 \label{fig:2_loop_daughters}
 \end{figure}

 As a simple example, consider the 2-loop diagram shown in
 Fig.~\ref{fig:2_loop_mother}.
 If  one of the lines is opened out so that it carries a
 thermal phase space factor as shown in the first of 
 Fig.~\ref{fig:2_loop_daughters}, the symmetry group 
 is reduced to the permutation of the two remaining lines so that
 the overall factor becomes $1/2!$.
 Originally, $q_G = 3!$, and now $q_S = 2!$.  The ratio
 $q_G/q_S = 3$ of course corresponds to the three choices we can make
 when selecting a line to replace.  
 
 If two lines from the diagram in
 Fig.~\ref{fig:2_loop_mother} are opened out and carry thermal weightings,
 we produce the second diagram in Fig.~\ref{fig:2_loop_daughters}.
 The symmetry group of the reduced diagram is the trivial identity. 
 Hence, the symmetry factor associated with the reduced diagram is 1.  
 However, since there are only ${}_3C_2 = 3$ choices of picking a pair of
 lines, the overall factor for the second diagram in 
 Fig.~\ref{fig:2_loop_daughters} is only $3/3! = 1/2!$.
 This is as it should be because the thermal lines have a reflection
 symmetry. 

 To make the connection between opened-up self-energy diagrams,
 such as those in Fig.~\ref{fig:2_loop_daughters},
 and the ${\cal T}$-matrices label the latter
 with 2 $k$'s, $m$ initial momenta $p_i$ and $m$ final momenta $q_i$.
 Eventually, we identify $p_i = q_i$
 and integrate over the thermal phase-space of all $p_i$'s 
 to make the contribution to the self-energy. 
 Suppose that all $p_i$ and $q_i$ belong to the same species.
 Then the matrix elements 
 \begin{eqnarray*} 
 & {\cal T}(k, p_1,\ldots, p_m; k, q_1,\ldots, q_m) &
 \non
 \inline{and}
 & {\cal T}(k, p_{\sigma_1},\ldots, p_{\sigma_m}; 
          k, q_{\sigma_1},\ldots, q_{\sigma_m})\;, &
 \end{eqnarray*}
 where $\sigma$ is one of the $m!$ permutation of $\{1,\ldots, m\}$,
 correspond to the same self-energy diagram when the $p_i$'s are 
 integrated over the thermal phase space.

 Even though there are $m!$ such permutations, the number of distinct
 ${\cal T}(k,\{p\};k,\{q\})$ need not equal $m!$. 
 For some permutations, the above two
 expressions may not only give the same contribution to the self-energy, 
 but actually be the same even before the integration.
 If we reconnect the $p_i$ vertices with the $q_i$ vertices,
 the resulting diagram is a self-energy diagram.
 Hence the number of permutations giving rise 
 to the same ${\cal T}$ must be equal to the order $q_{\rm t}$
 of the symmetry group $S_{\rm t}$ 
 of the reconnected part of the diagram.  
 Hence, the coefficient of this reconnected diagram is given by
 $m!/(q_{\rm t}\times q_{\rm z})$ where $q_{\rm z}$ is the symmetry
 factor associated with the ${\cal T}$. 
 
 Comparing this with $1/(q_{\rm t}\times q_{\rm z})$ which we had for the
 opened-up self-energy diagram,
 we see that a diagram
 generated from the self-energy is smaller by a factor of $m!$ than an
 actual scattering amplitude diagram.
 The generalization to the many species case is immediate.
 Thus if each particle species $i$ has $m_i$ external 
 thermally-weighted lines, the overall factor for the opened-up
 self-energy diagram is
 $1/S_{\{l_i^\sigma\}}~=~1/\prod_{i} m_i!$.


 \subsection{Disconnected Parts}
 \label{subsec:disconn}

 Suppose that, when the internal lines are opened out and given thermal
 weightings $\Gamma_\zeta$, we get a disconnected piece containing at
 least one internal vertex.  
 As long as this disconnected piece does not contain the largest time
 vertex, all cuts for this diagram,
 including the case with no cuts at all,
 are possible with the rest of the original diagram being kept common. 
 For the isolated subdiagram,
 this situation corresponds to having all the frequency denominators in
 Eq.~(\ref{eq:terms_in_disc}) and summing over all time orderings. 
 Consequently, the net contribution of such a disconnected part vanishes.
 In the real-time method this cancellation is referred to as 
 the vanishing of the sum of all circlings \cite{Semenoff}.

\begin{figure}[t]
 \begin{center}
 \leavevmode
 \epsfxsize = 12cm
 \epsfbox{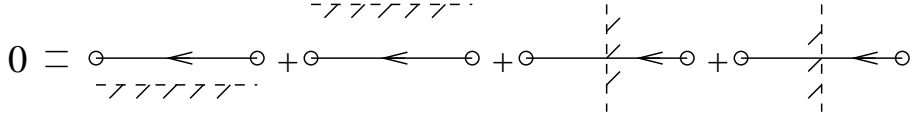}
 \end{center}
 \caption{Cancelling propagators.}
 \label{fig:prop_cancel}
 \end{figure}

 For an elementary example, consider the sum of all cut and uncut
 propagators depicted in Fig.~\ref{fig:prop_cancel}.
 The sum is
 \be
 (-i)^2 G_\zeta(k) + (i)^2 G^*_\zeta(k)
 +
 (-i)(i) \Delta^-_\zeta(k) 
 +
 (i)(-i) \Delta^+_\zeta(k) 
 =
 0\;,
 \ee
 where all the propagators are zero-temperature ones.
 From the structure of the
 propagators (\ref{eq:Del_B}) -- (\ref{eq:Fermion_G_star}), it is clear
 that the sum is always zero.


 \subsection{Self-Energy Insertions}
 \label{subsec:insertions} 

 For simplicity we only consider scalar particles without a chemical
 potential. 
 The argument presented here can be generalized immediately to other cases.
 Self-energy insertions are potentially hazardous when the
 propagators have poles at all four $k^0 = \pm E_k \pm i\epsilon$.
 This is the case for us due to the addition of the thermal phase space
 factor to the usual zero-temperature propagators
 ({\it c.f.} Eqs.~(\ref{eq:Del_B}) -- (\ref{eq:Fermion_G_star})). 
 The product of any two propagators with the same argument then contains
 pinching poles. 

 \begin{figure}[t]
 \begin{center}
 \leavevmode
 \epsfxsize=13cm
 \epsfbox{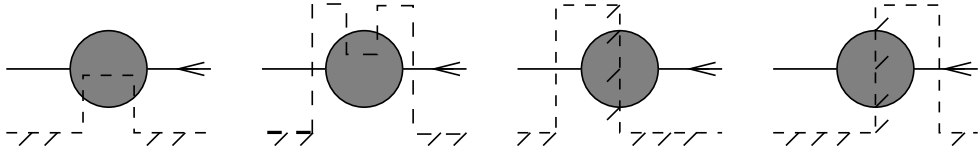}
 \end{center}
 \caption{Schematic depiction of the possible ways to cut a
 self-energy insertion diagram.}
 \label{fig:self_insertions}
 \end{figure}

 To show that in fact the pinching pole contributions all cancel, 
 consider the diagrams shown in Fig.~\ref{fig:self_insertions}. They
 differ only in the manner of cutting.
 Representing the blobs in the first and third diagrams
 by $C(k)$ and $L(k)$, respectively, the sum of the four
 diagrams can be written
 \be
 {\cal F}(k)
 & = &
 G_{\rm B}(k) C(k) G_{\rm B}(k) 
 + 
 \Delta_{\rm B}^-(k) C^*(k) \Delta_{\rm B}^+(k)
 \non
 & & {}
 -
 \Delta_{\rm B}^-(k) L(k) G_{\rm B}(k)
 -
 G_{\rm B}(k) L(-k) \Delta_{\rm B}^+(k)\;.
 \ee
 Extracting the coefficient of the pinching poles 
 $2(k^2-m^2+i\epsilon)^{-1}(k^2-m^2-i\epsilon)^{-1}$, we find
 \be
 {\cal F}_\scrpt{pinch}(k)
 & = &
 n_{\rm B}(k^0)N_{\rm B}(k^0) 
 \left( C(k) + C^*(k) \right)
 \non
 & & {}
 -
 n_{\rm B}(k_0)(1/2 + n_{\rm B}(k_0)) L(k)
 - 
 N_{\rm B}(k_0)(1/2 + n_{\rm B}(k_0)) L(-k)
 \;.
 \ee
 If ${\cal F}_\scrpt{pinch}(k) $ were non-zero,
 a self-energy insertion would cause an uncontrollable divergence.  
 Fortunately, ${\cal F}_\scrpt{pinch}(k)$ does vanish due to 
 the following properties of $C(k)$ and $L(k)$ 
 \cite{Semenoff,Kobes,Kobes90,jeon2}
 \be
 & C(k) + C^*(k) = L(k) + L(-k) &
 \label{eq:C_id}
 \\
 \inline{and}
 & L(-k) = e^{-k^0\beta}\,L(k) \;. &
 \label{eq:L_id}
 \ee
 The first identity (\ref{eq:C_id}) is the finite temperature version of 
 the optical theorem.  The second identity can be easily obtained by
 using $\Delta_{\rm B}^+(k) = e^{k^0\beta}\Delta_{\rm B}^-(k)$. 
 Therefore, pinching poles do not occur in self-energies insertions.

 The absence of pinching poles has two important consequences.  
 First, self-energy insertions do not cause uncontrollable divergences
 due to the cancellation between cut and uncut self-energies.  
 Second, {\em the propagators connected to a self-energy do not 
 produce thermal phase-space factors.}
 Without this second point it would not be possible to
 identify the coefficients of the expansion in
 the $\Gamma_\zeta$ factors with the scattering amplitudes.

 \subsection{Polarization Factors and the Overall Sign}
 \label{subsec:polar}

 To convert a Feynman diagram to a scattering amplitude, 
 one needs the polarization factors for the external lines.
 For an external gauge boson line, a polarization vector
 $\epsilon_\mu(p,s)$ orthogonal to the incoming momentum $p$ is needed.
 This is provided by the substitution
 \be
 g_{\mu\nu} 
 \to
 -\sum_{s=1,2} \epsilon_\mu(p,s)\epsilon_\nu^*(p,s) 
 \ee
 inside a Feynman diagram; this is valid due to the Ward identity. 
 In the Feynman gauge, the gauge boson
 phase space factor is proportional to the metric $g_{\mu\nu}$ which
 then provides appropriate $\epsilon_\mu$'s.

 For fermions, each incoming fermion (anti-fermion)
 line requires a factor of 
 the Dirac spinor $u_s(p)$ ($\bar{v}_s(p)$), 
 and each outgoing fermion (anti-fermion) line requires a
 factor of $\bar{u}_s(p)$ ($v_s(p)$). 
 For the external line corresponding to the self-energy momentum,
 one can use the Dirac spinor identity
 \be
 {\bf 1}
 = 
 {1\over 2m_F} \sum_s \left( u_s(k)\bar{u}_s(k) - v_s(k)\bar{v}_s(k) \right)
 \;,
 \ee
 where the spinors are normalized according to Peskin and Schroeder 
 \cite{peskin}.
 This yields 
 \be
 \Sigma_{\rm F}(k) 
 & = &
 {\bf 1} 
 \Sigma_{\rm F}(k) 
 {\bf 1}
 \non
 & = &
 {1\over 4 m_F^2} 
 \sum_s
 \left\{
 u_s(k) \Big[ \bar{u}_s(k)\Sigma_{\rm F}(k)u_s(k) \Big] \bar{u}_s(k)
 + 
 v_s(k) \Big[ \bar{v}_s(k)\Sigma_{\rm F}(k)v_s(k) \Big] \bar{v}_s(k)
 \right\}
 \;.
  \label{eq:fermion}
 \ee
 The factor 
 $\bar{u}_s(k)\Sigma_{\rm F}(k)u_s(k)/(2m_F)$
 then has a multiple scattering expansion in terms of spin 
 up-up and down-down scattering amplitudes involving the particles,
 while the factor $\bar{v}_s(k)\Sigma_{\rm F}(k)v_s(k)/(2m_F)$
 has a multiple scattering expansion in terms of spin 
 up-up and down-down scattering amplitudes involving the anti-particles.
 The mixed term vanishes since
 $\bar{u}(k)\gamma^0 v(k) = \bar{v}(k)\gamma^0 u(k) = 
 \bar{u}(k)k^\mu\gamma_\mu v(k) 
 = \bar{v}(k)k^\mu\gamma_\mu u(k) = 0$,
 and the self-energy, as well as the spectral density, must have the
 structure
 \be
 \Sigma_{\rm F}(k) = A(T, k)\gamma^\mu k_\mu - B(T, k) m_F 
 + C(T, k)\gamma^0 \;
 \ee
 at finite temperature.

 As an example, consider the nucleon self-energy 
 in a thermal pion medium.
 Considering only the strong interaction, we can regard both the pions
 and the nucleons as stable.
 The up-up component of the lowest order 
 nucleon self-energy can then be expressed as
 \be
 \Sigma_T^{++}(k)
 & \equiv &
 {\bar{u}_+(k)\Sigma_T(k) u_+(k) \over 2m_N}
 \non
 & = &
 -{1\over 2m_N}
 \sum_{a}
 \int {d^3 l_\pi \over (2\pi)^3 2E_\pi}\, n_B(E_\pi)\,
 {\cal T}^{++}_{N\pi_a\to N\pi_a}(k + l_\pi \to k + l_\pi)
 \non
 & = & 
 -2\pi \sum_{a}
 \int {d^3 l_\pi \over (2\pi)^3 E_\pi}\, n_B(E_\pi)\,
 \left( {\sqrt{s} \over m_N} \right)
 f^{++}_{N\pi_a}(k, l_\pi)
 \;,
 \label{eq:nucleon_self}
 \ee 
 where 
 ${\cal T}^{++}_{N\pi_a\to N\pi_a}(k + l_\pi \to k + l_\pi)$ is the sum
 of all scattering amplitudes including the cross terms. 
 In the last line we used ${\cal T} = 8\pi \sqrt{s}\, f$. 
 This expression differs by a factor of $\sqrt{s}/m_N$ from that
 given by Eletskii and Ioffe \cite{Ioffe}. At low temperature ($T \ll m_N$) 
 and in the nucleon rest frame, their expression may be justified because 
 the pion thermal energy is insignificant compared to the nucleon rest mass.

 For internal fermion lines which are opened up and given thermal
 weightings we use the relation
 \be
 &&{\hspace{-.3cm}}(k^\mu\gamma_\mu + m_F)\, 2\pi \, 
 \delta(k_0^2 - E_k^2)
 = 
 (k_+^\mu\gamma_\mu + m_F)\, {2\pi \over 2E_k} \, \delta(k_0 - E_k)
 -
 (k_-^\mu\gamma_\mu - m_F)\, {2\pi \over 2E_k} \, \delta(k_0 + E_k)
 \non
 & &{\hspace{2cm}}
 =\sum_s u_s(k_+) \bar{u}_s(k_+)\, {2\pi \over 2E_k} \, \delta(k_0 - E_k)
 -
 \sum_s v_s(k_-) \bar{v}_s(k_-)\, {2\pi \over 2E_k} \, \delta(k_0 + E_k)
 \;,
 \label{eq:rho_decomp}
 \ee
 where $k_\pm = (E_k, \pm {\bf k})$. Hence for both spin-1 gauge bosons 
 and spin-$\frac{1}{2}$ fermions the polarization factors are 
 all correctly accounted for.

 When fermions are involved, 
 our expansion must generate the correct overall sign of a diagram.
 The presence of a fermion loop in a diagram carries 
 an additional overall factor of $(-1)$. 
 When the fermion loop is broken this sign is carried by
 the thermal phase space factor, $\Gamma_F$ rather than 
 the scattering amplitude. The $v\bar{v}$ part of
 Eq.~(\ref{eq:rho_decomp}) carries an additional $(-1)$ due to the exchange
 of the fermion legs. 
 When a fermion line which is not a part of a fermion loop carries
 a thermal phase space factor, the $u\bar{u}$ term
 corresponds to a crossed diagram and the required factor of $(-1)$
 is provided by $\Gamma_F$.
 Hence, the overall signs are also correctly accounted for in our expansion.

 \section{Electron Self-Energy}
 \label{sec:elec_self}

 \begin{figure}[t]
 \begin{center}
 \leavevmode
 \epsfxsize=6cm
 \epsfbox{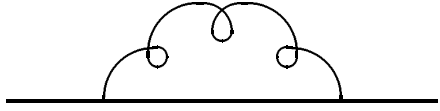}
 \end{center}
 \caption{The one-loop electron self-energy. The curly line represents
 the photon.}
 \label{fig:elec_self}
 \end{figure}

 As an example of applying the method developed here, consider the
 electron self-energy at temperatures $T \ll m_e$ obtained from the diagram
 of Fig.~\ref{fig:elec_self}.  At these low
 temperatures, we can neglect the electron thermal correction because
 it is suppressed by a factor of $e^{-\beta m_e}$.
 The self-energy written down for the nucleon (\ref{eq:nucleon_self}) is 
 valid for the electron if the pion is changed to a photon with the
 summation referring to the photon polarization:
 \be
 \Sigma_T^{++}(p)
 & = &
 -{1\over 2m_e}
 \sum_{s=1,2}
 \int {d^3 k \over (2\pi)^3 2|{\bf k}|}\, n_B(|{\bf k}|)\,
 {\cal T}^{++}_{e\gamma\to e\gamma}(p + k\to p + k)
 \label{eq:electron_self}
 \;.
 \ee 
 After the photon polarization summation,
 the spin up-up component of the 
 tree-level Compton scattering amplitude is given by
 ({\it e.g.} Ref. \cite{peskin})
 \be
 {\cal T}^{++}_{e\gamma\to e\gamma}(p + k\to p + k)
 & = &
 e^2 \bar{u}_+(p)
 \left[
 {\gamma^\mu\rlap{/}k \gamma_\mu+2\rlap{/}p\over 2p{\cdot}k + i\epsilon}
 +
 {-\gamma^\mu\rlap{/}k \gamma_\mu+2\rlap{/}p\over -2p{\cdot}k + i\epsilon}
 \right]
 u_+(p)
 \;.
 \ee
 In the electron rest frame, this reduces to  
 \be
 {\cal T}^{++}_{e\gamma\to e\gamma}(p + k\to p + k)
 = 
 -4 e^2
 \left[
 1 +\frac{2m_e^2\epsilon i}{4m_e^2|{\bf k}|^2+\epsilon^2}
 \right]
 \;.
 \ee
 The real part in the rest frame is then,
 \be
 {\rm Re}\,\Sigma_T^{++}(m_e)
 & = &
 {2 e^2\over m_e}
 \int {d^3 k \over (2\pi)^3 2|{\bf k}|}\, n_B(|{\bf k}|)\,
 \non
 & = &
 {e^2 T^2\over 12 m_e} 
 = 
 {\pi\alpha T^2 \over 3 m_e}
 \;.
 \ee 
 The result, of course, coincides with 
 a previous calculation \cite{donoghue}. 
 The imaginary part is 
 \be
 {\rm Im}\,\Sigma_T^{++}(m_e)
 & = & \lim_{\epsilon\to0}
 4e^2 m_e\epsilon
 \int {d^3 k \over (2\pi)^3 2|{\bf k}|}\, \frac{n_B(|{\bf k}|)}
 {(4m_e^2|{\bf k}|^2+\epsilon^2)}
 \non
 & = &
 {e^2 T \over 4\pi} = \alpha T
 \;.
 \ee 
 This agrees with the leading term given by Henning {\it et al.}
 \cite{henning}.

 \section{Conclusion}
 \label{sec:concl}

 In this paper, the multiple scattering expansion of the thermal correction
 to the retarded
 self-energy is presented starting from the imaginary-time formalism.
 The leading order term of this expansion corresponds to
 the often used rule that the thermal correction to the self-energy
 is the thermal phase-space times 
 the scattering amplitude.
 Although the formal expansion can be always made, 
 we have argued that the expansion is useful only if the minimum of 
 $|m_a \pm \mu_a|$ is large compared to the temperature.
 We have also demonstrated the connection between the self-energy and the 
 thermal part of the grand potential. 

 The result presented here may be used in two ways.  
 One is to use existing calculations of scattering amplitudes to
 calculate the self-energy.  In this way, the considerable effort usually
 needed to evaluate thermal correlation functions can be much reduced. 
 The other way is simply to use experimental scattering amplitudes to
 calculate the self-energy.  For interactions involving large coupling
 constants, this may be the only reliable way to calculate the
 thermal correction to the self-energy. 

 The reliable calculation of medium effects is important in analyzing data
 from heavy-ion collision experiments.  
 For instance, 
 how the $\rho$-meson behaves in-medium can greatly influence the
 dilepton spectrum in heavy ion collisions \cite{Shuryak,Ko}.
 The in-medium effect becomes even more important 
 in future RHIC (Relativistic Heavy Ion Collider) experiments.  
 The main goal of RHIC is to find the quark-gluon plasma.  
 Hence, it is crucial to understand the hadronic part of the in-medium
 finite-temperature effect so as to separate this signal from that of the
 long-sought quark-gluon plasma.
 Even though a full theory of hadron scattering is lacking, there is a 
 considerable amount of data on scattering cross-sections accumulated in
 the past decades.  Utilizing those data we may at least
 phenomenologically separate a truly hadronic effect from the effect 
 of a new state of matter.

 \section*{Acknowledgment}

  We would like to thank J.I.~Kapusta for encouragement and
 enlightening discussion.  We also would like to thank 
 H.-B.~Tang for helpful suggestions.  P.J.E. also thanks
 M. Prakash and P. Lichard for stimulating interactions.
 This work was supported by the US Department of Energy
 under grant DE-FG02-87ER40328.

 \appendix

 \section{Euclidean Propagators}
 \label{app:eucl}
 
 \subsection{Bosonic Propagator With Chemical Potential}
 
 When the chemical potential is non-zero, we must at least deal with a
 complex field. 
 The effective Hamiltonian is 
 \be
 \hat{K} \equiv \hat{H} - \sum_a \mu_a \hat{Q}_a 
 \;.
 \ee
 For notational convenience we suppress spatial indices in this
 section.

 The spectral density for the propagator is defined to be \cite{Fetter}
 \be
 \rho_{\rm B}(\omega)
 & = & 
 \int dt\, e^{i\omega t}\,
 \langle [\phi(t), \phi^\dagger] \rangle
 \non
 & = &
 \sum_{m,n}
 \int dt\, e^{i\omega t}\,
 \left(
 e^{-\beta K_m} e^{it(K_m - K_n)} \phi_{mn}\phi^\dagger_{nm} 
 -
 e^{-\beta K_n} \phi^\dagger_{nm} e^{it(K_m - K_n)} \phi_{mn} 
 \right)
 \non
 & = &
 \sum_{m,n}
 2\pi \delta(\omega + K_m - K_n)\,
 \left( e^{-\beta K_m} - e^{-\beta K_n} \right) 
 \phi_{mn}\phi^\dagger_{nm}
 \non
 & = &
 \left( 1 - e^{-\beta \omega} \right) 
 \sum_{m,n}
 2\pi \delta(\omega + K_m - K_n)\,
 \phi_{mn}\phi^\dagger_{nm}
 e^{-\beta K_m}\;.
 \ee
 The Euclidean propagator is given by
 \be
 G_{\rm B}(\tau) 
 & = &
 \langle {\cal T}\phi(\tau)\phi^\dagger(0) \rangle
 \non
 & = &
 \theta(\tau) 
 {\rm Tr}\left( e^{-\beta \hat{K}}e^{\tau \hat{K}}
                \phi e^{-\tau \hat{K}} \phi^\dagger\right)
 +
 \theta(-\tau) 
 {\rm Tr}\left( e^{-\beta \hat{K}}\phi^\dagger e^{\tau \hat{K}}
                \phi e^{-\tau \hat{K}} \right)
 \non
 & = &
 \int \dom{\omega}\,
 \left(
 e^{-\omega \tau}
 [1 + n_{\rm B}(\omega)]\,\rho_B(\omega) 
 \theta(\tau) 
 +
 e^{-\omega \tau}
 n_{\rm B}(\omega)\,\rho_B(\omega) 
 \theta(-\tau) 
 \right)
 \non
 & = &
 \int \dom{\omega}\,
 [1 + n_{\rm B}(\omega)]
 \left(
 e^{-\omega \tau} \,\rho_{\rm B}^+(\omega)\, \theta(\tau) 
 +
 e^{\omega \tau} \,\rho_{\rm B}^-(\omega)\, \theta(-\tau) 
 \right)\;,
 \ee
 where we used
 \begin{equation}
 n_{\rm B}(-\omega) = -[1 + n_{\rm B}(\omega)]\;,
 \end{equation}
 and defined
 \begin{equation}
 \rho_{\rm B}^+(\omega) \equiv \rho_{\rm B}(\omega)
 \quad;\quad 
 \rho_{\rm B}^-(\omega) \equiv -\rho_{\rm B}(-\omega)
 \;.
 \end{equation}

 To find $\rho^+_{\rm B}(\omega)$, use \cite{Fetter}
 \be
 G_{\rm B}(i\nu_B, {\bf k})
 =
 {1\over E_k^2 - (i\nu_B + \mu)^2}
 =
 \int \dom{\omega}\,
 {\rho^+_{\rm B}(\omega, {\bf k}) \over \omega - i\nu_B}\;,
 \ee
 with $E_k=\sqrt{{\bf k}^2+m^2}$, and take the discontinuity across the real 
 axis to get
 \be 
 \rho^+_{\rm B}(\omega, {\bf k})
 & = &
 {\rm sign}(\omega + \mu)\,
 2\pi\,\delta\left( (\omega + \mu)^2 - E_k^2 \right)
 \\
 \noalign{\hbox{and}}
 \rho^-_{\rm B}(\omega, {\bf k})
 & = & 
 -\rho^+_{\rm B}(-\omega, {\bf k})
 = 
 {\rm sign}(\omega - \mu)\,
 2\pi\,\delta\left( (\omega - \mu)^2 - E_k^2 \right)
 \;.
 \ee
 
 \subsection{Fermionic Propagators with Chemical Potential}

 The spectral density for the fermion propagator is defined to be
 \be
 \rho_{\rm F}(\omega)
 & = &
 \int dt\, e^{i\omega t}\,
 \langle \{ \psi(t), \bar{\psi} \} \rangle
 \non
 & = &
 \sum_{m,n}
 2\pi \delta(\omega + K_m - K_n)\,
 \left( e^{-\beta K_m} + e^{-\beta K_n} \right) 
 \psi_{mn}\bar{\psi}_{nm}
 \non
 & = &
 \left( 1 + e^{-\beta\omega} \right) 
 \sum_{m,n}
 2\pi \delta(\omega + K_m - K_n)\,
 \psi_{mn}\bar{\psi}_{nm}
 e^{-\beta K_m}\;,
 \ee
 where the brace denotes an anticommutator and we used the fact that 
 matrix elements such as $\psi_{mn}$ are c-numbers.

 The Euclidean propagator is given by
 \be
 G_{\rm F}(\tau) 
 & = &
 \langle {\cal T}\psi(\tau)\bar\psi(0) \rangle
 \non
 & = &
 \theta(\tau) 
 {\rm Tr}\left( e^{-\beta \hat{K}}e^{\tau \hat{K}}
                \psi e^{-\tau \hat{K}} \bar{\psi}\right)
 -
 \theta(-\tau) 
 {\rm Tr}\left( e^{-\beta \hat{K}}\bar{\psi} e^{\tau \hat{K}}
                \psi e^{-\tau \hat{K}} \right)
 \non
 & = &
 \int \dom{\omega}\,
 \left(
 e^{-\omega \tau}
 [1 - n_{\rm F}(\omega)]\,\rho^+_{\rm F}(\omega) 
 \theta(\tau) 
 -
 e^{-\omega \tau}
 n_{\rm F}(\omega)\,\rho^+_{\rm F}(\omega) 
 \theta(-\tau) 
 \right)
 \non
 & = &
 \int \dom{\omega}\,
 [1 - n_{\rm F}(\omega)] 
 \left(
 e^{-\omega \tau}\,\rho_{\rm F}^+(\omega) 
 \theta(\tau) 
 + 
 e^{\omega \tau}\,\rho_{\rm F}^-(\omega) 
 \theta(-\tau) 
 \right)\;,
 \ee
 where the change of variable $\omega \to -\omega$ was made for the
 second term and we used
 \be
 n_{\rm F}(-\omega) 
 = {1\over e^{-\omega\beta} + 1}
 = 1 - n_{\rm F}(\omega)\;.
 \ee
 Also, we defined
 \be
 \rho_{\rm F}^+(\omega) \equiv \rho_{\rm F}(\omega)\quad;\quad
 \rho_{\rm F}^-(\omega) \equiv -\rho_{\rm F}(-\omega)\;.
 \ee 

 Hence, for both bosons and fermions we have
 \be
 G_\zeta(\tau) 
 = 
 \int \dom{\omega}\,
 N_\zeta(\omega)
 \left(
 e^{-\omega \tau}\,\rho_\zeta^+(\omega) 
 \theta(\tau) 
 + 
 e^{\omega \tau}\,\rho_\zeta^-(\omega) 
 \theta(-\tau) 
 \right)\;,\label{eq:gbf}
 \ee
 with
 \be
 \rho_\zeta^+(\omega)
 & = & 
 -\rho_\zeta^-(-\omega) \;,
 \\
 \inline{and}
 N_\zeta(\omega) 
 &=& 
 1 + (-1)^\zeta n_\zeta(\omega)
 =
 \theta(\omega) + (-1)^\zeta {\rm 
 sign}(\omega)n_\zeta(|\omega|)\;,\label{eq:bign}
 \ee
 where
 \be
 (-1)^{\rm B} \equiv 1 
 \ \  \hbox{and}\ \ 
 (-1)^{\rm F} \equiv -1 
 \;.
 \ee

 To find $\rho^+_{\rm F}(\omega)$, start from the Euclidean propagator
 \be
 G_{\rm F}(\nu_{\rm F}, {\bf p})
 = 
 { i\left(\nu_{\rm F} - i\mu \right)\gamma^0 - {\bf p}{\cdot}\bgamma + m
 \over 
 (\nu_{\rm F} - i\mu)^2 + E_p^2}
 =\int \dom{\omega}\,
 {\rho^+_{\rm F}(\omega, {\bf k}) \over \omega - i\nu_F}\;,
 \ee
 with $E_p=\sqrt{{\bf p}^2+m^2}$, and take the discontinuity across the 
 real axis to get
 \be
 \rho_{\rm F}^+(\omega, {\bf p})
 =
 \left( (\omega + \mu)\gamma^0 - {\bf p}{\cdot}\bgamma + m \right)
 {\rm sign}(\omega+\mu)\,2\pi\delta((\omega+\mu)^2 - E_p^2)
 \;.
 \ee

 \section{Real Time Propagators} 
 \label{app:real}

 For both bosons and fermions the real time version of Eq. (\ref{eq:gbf}) is 
 \be
 G_\zeta(t) 
 & = &
 \int \dom{\omega}\,
 N_\zeta(\omega)
 \left(
 e^{-i\omega t} \,\rho_\zeta^+(\omega)\, \theta(t) 
 +
 e^{i\omega t} \,\rho_\zeta^-(\omega)\, \theta(-t) 
 \right)
 \;.
 \ee
 The Fourier transform yields the momentum space propagator
 \be
 G_\zeta(k) 
 & = &
 \int \dom{\omega}\,
 N_\zeta(\omega)
 \left(
 \int_0^\infty dt\,
 e^{ik^0t -\epsilon t}
 e^{-i\omega t}\,
 \rho_\zeta^+(\omega) 
 +
 \int_{-\infty}^0 dt\,
 e^{ik^0t +\epsilon t}
 e^{i\omega t}\,
 \rho_\zeta^-(\omega)
 \right)
 \non
 & = & 
 \int \dom{\omega}\,
 [\theta(\omega) + (-1)^\zeta {\rm sign}(\omega)n_\zeta(|\omega|)]
 \left(
 {\rho_\zeta^+(\omega) \over \epsilon + i(\omega - k^0)} 
 +
 {\rho_\zeta^-(\omega) \over \epsilon + i(\omega + k^0)} 
 \right)\;,
 \ee 
 using Eq. (\ref{eq:bign}). The propagator has two terms.  
 The first corresponds to the zero temperature case
 \be
 G_\zeta^0(k)
 =
 \int \dom{\omega}\,
 \theta(\omega)
 \left(
 {\rho_\zeta^+(\omega) \over \epsilon + i(\omega - k^0)} 
 +
 {\rho_\zeta^-(\omega) \over \epsilon + i(\omega + k^0)} 
 \right)\;,
 \ee
 which yields the standard Minkowski propagators given in the text.
 The second term corresponds to the finite temperature phase space 
 factor
 \be
 \Gamma_\zeta(k) 
 & = &
 (-1)^\zeta 
 \int \dom{\omega}\,
 {\rm sign}(\omega)n_\zeta(|\omega|)
 \left(
 {\rho_\zeta^+(\omega) \over \epsilon + i(\omega - k^0)} 
 +
 {\rho_\zeta^-(\omega) \over \epsilon + i(\omega + k^0)} 
 \right)
 \non
 & = &
 (-1)^\zeta 
 {\rm sign}(k^0)n_\zeta(|k^0|)
 \rho_\zeta^+(k) \;,
 \ee 
 where we used $\rho_\zeta^-(\omega) = -\rho_\zeta^+(-\omega)$.
 
 Cut propagators can also be decomposed into zero and non-zero
 temperature parts.  The cut propagator for a particle is 
 \begin{equation}
 \Delta_\zeta^+(k)\equiv N_\zeta(k^0)\rho_\zeta^+(k)
 =  \theta(k^0)\rho_\zeta^+(k) +
 (-1)^\zeta {\rm sign}(k^0)n_\zeta(|k^0|)\, \rho_\zeta^+(k)
 \;.
 \end{equation}
 Letting the momentum and frequency follow the charge,
 we get the cut propagator for an anti-particle
 $N_\zeta(k^0)\rho_\zeta^-(k) \to N_\zeta(-k^0)\rho_\zeta^-(-k)$ which
 is
 \be
 \Delta_\zeta^-(k)\equiv N_\zeta(-k^0)\rho_\zeta^-(-k)
 & = & 
 (-1)^\zeta\, n_\zeta(k^0)\, \rho_\zeta^+(k)
 \non
 & = &
 -\theta(-k^0)\rho_\zeta^+(k)
 +
 (-1)^\zeta\, {\rm sign}(k^0)\, n_\zeta(|k^0|)\, \rho_\zeta^+(k)\;,
 \ee 
 using
 \be
 n_\zeta(k^0) = -(-1)^\zeta \theta(-k^0) + {\rm sign}(k^0)n_\zeta(|k^0|)
 \;.
 \ee
 We also note that 
 $\Delta_\zeta^+(k)=(-1)^\zeta e^{\beta k_0}\Delta_\zeta^-(k)$.

\section{Derivative Couplings}
\label{app:derivative}

  For the time derivative of the propagator in Eq. (\ref{eq:gbf}) we have 
 \be
 {\partial \over \partial \tau}G_\zeta(\tau, {\bf k})
 & = &
 \int \dom{\omega}\, N_{\zeta}(\omega)\,
 \left( 
 -\omega \rho_{\zeta}^+(\omega, {\bf k})\,e^{-\omega \tau}\,\theta(\tau)\,
 +\omega \rho_{\zeta}^-(\omega, {\bf k})\,e^{\omega \tau}\,\theta(-\tau)\,
 \right) 
 \non
 & & {}
 +
 \int \dom{\omega}\, N_{\zeta}(\omega)\,
 \left( 
 \rho_{\zeta}^+(\omega, {\bf k})
 - 
 \rho_{\zeta}^-(\omega, {\bf k})
 \right) 
 \delta(\tau)
 \non
 & = &
 \int \dom{\omega}\, N_{\zeta}(\omega)\,
 \left( 
 -\omega \rho_{\zeta}^+(\omega, {\bf k})\,e^{-\omega \tau}\,\theta(\tau)\,
 +\omega \rho_{\zeta}^-(\omega, {\bf k})\,e^{\omega \tau}\,\theta(-\tau)\,
 \right) 
 \non & & {}
 +
 \int \dom{\omega}\,\rho_{\zeta}^+(\omega, {\bf k})\delta(\tau)
 \;.
 \ee 
 Thus the spectral density for the time derivative of the propagator is
 $ -\omega\rho_{\zeta}^{+}(\omega, {\bf k})$ and there is also 
 a $\delta(\tau)$ term. The latter vanishes for bosons since 
 $\int d\omega\, \rho_{\rm B}^{+}(\omega, {\bf k}) = 0$.
 In general, we can say
 \begin{equation}
 \partial_\tau^r G_\zeta(\tau, {\bf k})
 =
 \tilde{G}_\xi(\tau, {\bf k}) 
 +
 \sum_{l=0}^{r-1}
 \delta^{(l)}(\tau) F_l({\bf k})\;, \label{eq:gderiv}
 \end{equation}
 where the label $\xi$ includes both $\zeta$ and the number of derivatives $r$
 so that $\tilde{G}_\xi$
 has the spectral density $(-\omega)^r \rho_\zeta^+$. In the second term 
 of (\ref{eq:gderiv}) 
 $\delta^{(l)}(\tau) = \partial_\tau^l\delta(\tau)$ and the
 sum does not contribute for $r=0$.
 Then, after all the $\delta$-functions in Eq. (\ref{eq:gderiv}) are
 integrated over, we have 
 \be
 \lefteqn{
 C_{N+1}^{(\Gamma)}(\{ {\bf q}_l, i\nu_l\})
 } & &
 \non
 & = &
 \int_0^\beta \prod_{i=0}^V d\tau_i 
 \exp\left( i\sum_{l=0}^N \nu_l\tau_l \right)\,
 \int\prod_{L\in \Gamma} {d^3 k_L\over(2\pi)^3}\,
 A_0\left(\{ {\bf k}_\alpha \} \right)\,
 \prod_{\alpha \in \Gamma}{}
 \tilde{G}_{\xi_\alpha}(\tau^\alpha_a - \tau^\alpha_b)
 \non
 & & {}
 +
 \sum_r
 \int_0^\beta \prod_{i\ne r}^V d\tau_i 
 \exp\left( i\sum_{l\ne r}^N \bar{\nu}_l\tau_l \right)\,
 \int\prod_{L\in \Gamma} {d^3 k_L\over(2\pi)^3}\,
 A_1\left(\{ {\bf k}_\alpha, i\nu_r \} \right)\,
 \prod_{\alpha \in \Gamma'}{}
 \tilde{G}_{\xi_\alpha}(\tau^\alpha_a - \tau^\alpha_b)
 \non
 & & {}
 +
 \sum_{q,r}
 \int_0^\beta \prod_{i\ne r,q}^V d\tau_i 
 \exp\left( i\sum_{l\ne r,q}^N \bar{\bar{\nu}}_l\tau_l \right)\,
 \int\prod_{L\in \Gamma} {d^3 k_L\over(2\pi)^3}\,
 A_2\left(\{ {\bf k}_\alpha, i\nu_r, i\nu_q \} \right)\,
 \prod_{\alpha \in \Gamma''}{}
 \tilde{G}_{\xi_\alpha}(\tau^\alpha_a - \tau^\alpha_b)
 \non
 & & {}
 + \cdots
 \;,
 \label{eq:C_Gamma_2}
 \ee
 where in $\Gamma'$ a pair of vertices refer to the same time, in
 $\Gamma''$ two pairs of vertices refer to the same times, and so on.
 The bars over the $\nu_l$ indicates that some of them are now
 combinations of the original $\nu_l$'s.
 
 If a total of $n$ derivatives is contained in the expression for the 
 diagram $\Gamma$, then performing the time integrals we have 
 \be
 {C}_{N{+}1}^{(\Gamma)} (\{ {\bf q}_l, i\nu_l\})
 =
 \sum_{r=0}^{n}
 \sum_{{\Gamma_{\sigma}^r} \subset \Gamma_r} \;
 C_{N+1}^{(\Gamma_\sigma^r)}(\{ {\bf q}_l, i\nu_l\})
 \ee 
 where 
 \be
 \lefteqn{C_{N+1}^{(\Gamma_\sigma^r)}(\{ {\bf q}_l, i\nu_l\})}
 \non
 & = &
 \int
 \prod_{L \in \Gamma} {d^3 k_L \over (2\pi)^3} \,
 \prod_{\alpha \in \Gamma_r}
 \left(
 \int {d\omega_{\alpha} \over 2\pi} \, 
 N_{\zeta_\alpha}(\omega_\alpha)\,
 \rho_{\xi_\alpha}^{s_\sigma}(k_\alpha)
 \right)\,
 A_r\left(\{ {\bf k}_\alpha \}, \{ i\nu_l \} \right)\,
 \prod_{{
 \hbox{\scriptsize{\rm intervals}}
 \atop
 \hbox{$\scriptstyle{V-r \geq j \geq 1}$} }}
 \Bigl(
 \Lambda_j^{\sigma} -  i\nu_j^{\sigma}
 \Bigr)^{-1}
 \;.
 \non
 \label{eq:C_Gamma_3}
 \ee
 We know that in the $\beta\to \infty$ limit,
 the frequency denominators we get from
 Eq.~(\ref{eq:C_Gamma_3}) 
 must add up to make the Wick rotated $\partial_t^r G_\zeta(t)$.
 That is, the coefficients
 $A_r$ must be such that if we change $N\to \theta$,
 $i\nu_l \to q_l^0$ and add $-i\epsilon$ to
 $\Lambda_j^\sigma$ in Eq. (\ref{eq:C_Gamma_3}), we get the zero 
 temperature $N+1$-point function:
 \begin{equation}
 {C}_{N{+}1}^{(\Gamma)} (\{ {\bf q}_l, i\nu_l\})
 \rightarrow {D}_{N{+}1}^{(\Gamma)}(\{q_l\}) =i^V
 \int
 \prod_{L \in \Gamma}
 {d^4 k_{L} \over (2\pi)^4} \,
 A(\{k_L\})
 \prod_{\alpha \in \Gamma}
 G_{\zeta_\alpha} (k_{\alpha}) \;,
 \end{equation}
 as in Eq. (\ref{eq:no_cut}). Arguments similar to those in 
 Subsec. \ref{sec:retarded} can be 
 applied to cut diagrams so that summing the time-ordered diagrams on the
 unshaded side leads to the usual Feynman rules, while for the shaded side
 complex conjugate Feynman rules apply. Thus the multiple scattering 
 expansion of the self-energy in Sec. 3 is also applicable when derivative 
 couplings are present.

 \section{The Thermodynamic Potential}
 \label{app:omega}

 Here we consider the multiple scattering expansion of the
 thermodynamic grand potential defined by
 \be
 {\Omega\over V} \equiv -{1\over \beta V}\,\ln Z
 \ee
 where $Z$ is the partition function and $V$ is the volume.
 In the imaginary-time formalism, the thermodynamic potential is the sum
 of all connected ``vacuum'' graphs.  
 Analytic continuation of such a result may at first appear to
 be a poorly defined concept since there are no external frequencies to
 start with. 
 Nevertheless, it is possible to consider the vacuum graphs as the zero
 frequency limit of the $N$-point functions \cite{EvansZero} and, further,
 it is sufficient to consider {\em retarded} correlation functions due
 to the reality of the thermodynamic potential.

\begin{figure}[t]
 \begin{center}
 \leavevmode
 \epsfysize=6cm
 \epsfbox{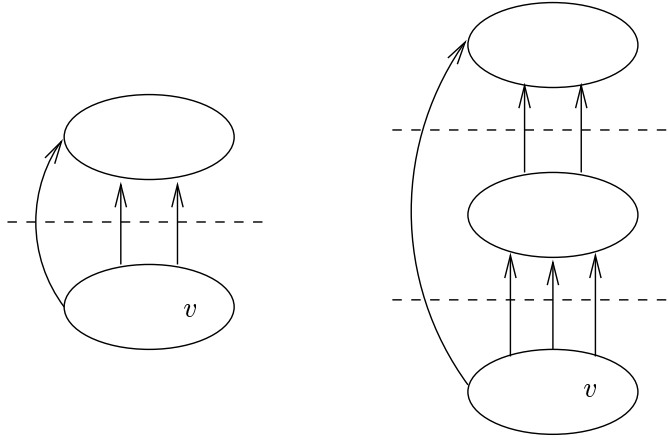}
 \end{center}
 \caption{Diagrams for the thermodynamic potential. The dashed line 
 indicates a cut.}
 \label{fig:many_cuts}
 \end{figure}

 To calculate $\Omega$, we regard the $\hat{\varphi}$ in 
 Eq.~(\ref{eq:retn1}) as ``external" interaction vertices which will 
 contain several 
 fields. The retarded functions then consist entirely of all possible 
 interaction vertices and there are zero external frequencies and
 momenta entering or leaving the diagram. Eq.~(\ref{eq:R_before_sum}) can 
 then be written
 \be
 {R}_{N{+}1}^{(\Gamma_v)} (0)
 & = & \displaystyle
 \sum_{{\Gamma_{\sigma}} \subset \Gamma} \;
 \int
 \prod_{L \in \Gamma} {d^3 k_L \over (2\pi)^3} \,
 \prod_{\alpha \in \Gamma}
 \left(
 \int {d\omega_{\alpha} \over 2\pi} \, 
 N_{\zeta_\alpha}(\omega_\alpha)\,
 \rho_{\zeta_\alpha}^{s_\sigma}(k_\alpha)
 \right)
 A_{V+1}^{N+1}
 \non
 & & \qquad {} \times
 \prod_{
 {
 \hbox{\scriptsize{\rm intervals}}
 \atop
 \hbox{$\scriptstyle{V \geq j \geq v+1}$}
 }
 }
 \Bigl(
 \Lambda_j^{\sigma} + i\epsilon
 \Bigr)^{-1}
 \prod_{
 {
 \hbox{\scriptsize{\rm intervals}}
 \atop
 \hbox{$\scriptstyle{v \geq j' \geq 1}$}
 }
 }
 \Bigl(
 \Lambda_{j'}^{\sigma} - i\epsilon
 \Bigr)^{-1}
 \;.
 \ee
 Here the fixed vertex $v$ is arbitrarily chosen for each diagram.
 Instead of the identity (\ref{eq:terms_in_disc}),
 we use 
 \be
 {1\over \Lambda_j^{\sigma} + i\epsilon}
 =
 {1\over \Lambda_j^{\sigma} - i\epsilon}
 -
 2\pi i \delta(\Lambda_j^\sigma)
 \;,
 \ee
 to get
 \be
 {R}_{N{+}1}^{(\Gamma_v)} (0)
 & = & \displaystyle
 \sum_{{\Gamma_{\sigma}} \subset \Gamma} \;
 \int
 \prod_{L \in \Gamma} {d^3 k_L \over (2\pi)^3} \,
 \prod_{\alpha \in \Gamma}
 \left(
 \int {d\omega_{\alpha} \over 2\pi} \, 
 N_{\zeta_\alpha}(\omega_\alpha)\,
 \rho_{\zeta_\alpha}^{s_\sigma}(k_\alpha)
 \right)
 A_{V+1}^{N+1}
 \non
 & & \qquad {} \times
 \Bigg[
 \prod_{
 {
 \hbox{\scriptsize{\rm intervals}}
 \atop
 \hbox{$\scriptstyle{V \geq j' \geq 1}$}
 }
 }
 \Bigl(
 \Lambda_{j'}^{\sigma} - i\epsilon
 \Bigr)^{-1}
 \non & & \qquad\quad {} 
 +(-i) 
 \sum_{j = v+1}^V\,
 2\pi\delta( \Lambda_j^{\sigma} )
 \prod_{
 {
 \hbox{$\scriptstyle{V \geq k \geq 1}$}
 \atop
 \hbox{$\scriptstyle{k \ne j}$}
 }
 }
 \Bigl(
 \Lambda_{k}^{\sigma} - i\epsilon
 \Bigr)^{-1}
 \non & & \qquad\quad {} 
 +
 (-i)^2 
 \sum_{j,k = v+1}^V\,
 2\pi\delta(\Lambda_j^{\sigma})
 2\pi\delta(\Lambda_k^{\sigma})
 \prod_{
 {
 \hbox{$\scriptstyle{V \geq l \geq 1}$}
 \atop
 \hbox{$\scriptstyle{l \ne j,k}$}
 }
 }
 \Bigl(
 \Lambda_{k}^{\sigma} - i\epsilon
 \Bigr)^{-1}
 \non & & \qquad \quad {}
 + \ldots
 +
 (-i)^{V-v} 
 \prod_{
 {
 \hbox{\scriptsize{\rm intervals}}
 \atop
 \hbox{$\scriptstyle{V \geq j \geq v+1}$}
 }
 }
 2\pi\delta(\Lambda_{j}^{\sigma})
 \prod_{
 {
 \hbox{$\scriptstyle{v \geq k \geq 1}$}
 }
 }
 \Bigl(
 \Lambda_{k}^{\sigma} - i\epsilon
 \Bigr)^{-1}
 \Bigg]
 \;.
 \label{eq:Omega_before_sum}
 \ee
 In this way the denominators contain only $-i\epsilon$ and consequently,
 after summing over all time orderings $\sigma$, the 
 result can be expressed entirely in terms of $G_\zeta(k)$ and 
 $\Delta_\zeta^{\pm}(k)$, {\it i.e.}
 the complex conjugate propagator $G_\zeta^*(k)$ does not occur.
 The price paid is the appearance of multiple cuts since
 each of the $\delta$-functions in Eq.~(\ref{eq:Omega_before_sum})
 represents a cut. So the result of the time ordering summation will
 yield uncut diagrams plus those with a sequence of cuts
 illustrated in Fig.~\ref{fig:many_cuts}. 
 The blobs in these diagrams involve the propagators $G_\zeta(k)$, while the cut
 lines require $\Delta_\zeta^{\pm}$ according to the direction of the
 momentum, as before.  

 Following the same procedure as before, we can now expand the diagrams
 in the number of thermal phase space factors $\Gamma_\zeta$. 
 The contribution of the first term in Eq.~(\ref{eq:Omega_before_sum}) 
 to the thermal part of $\Omega$ is  
 \be
 {\Omega_T^{(1)}\over V}
 = -
 \sum_{n \ge 2, \sigma}
 {1\over S_{\{l_i^\sigma \}}}
 \int \prod_{i=1}^{n}d\Gamma_i^\sigma\,
 \langle \{l_i^\sigma \} |  {\cal T} | \{l_i^\sigma \} \rangle_\scrpt{conn}
 \;,
 \ee
 where we have excluded diagrams with a single thermal weighting since we
 assume that the physical masses of the particles are used 
 in the propagators.
 The contribution of diagrams with $(m-1)$ cuts is 
 \be
 {\Omega_T^{(m)}\over V}
 = - \sum_{n \ge 2, \sigma}
 {1\over S_{\{l_i^\sigma \}}}
 \int \prod_{i=1}^{n}d\Gamma_i^\sigma\,
 \langle \{l_i^\sigma \} |  
 {\cal T}_v (-i{\cal T})^{m-1} 
 | \{l_i^\sigma \} \rangle_\scrpt{conn}
 \;.
 \ee
 Here ${\cal T}_v$ contains the arbitarily-chosen fixed vertex, $v$. 
 If, however, we remove this restriction and allow $v$ to lie in any of the
 ${\cal T}$ matrices, while compensating for the multiple counting,
 we obtain
 \be
 {\Omega_T^{(m)}\over V}
 =
 -i
 \sum_{n \ge 2, \sigma}
 {1\over S_{\{l_i^\sigma \}}}
 \int \prod_{i=1}^{n}d\Gamma_i^\sigma\,
 {1\over m}
 \langle \{l_i^\sigma \} |  
 (-i{\cal T})^m 
 | \{l_i^\sigma \} \rangle_\scrpt{conn}
 \;.
 \ee
 
 Summing over the number of cuts, we then have
 \be
 {\Omega_T\over V}
 =
 \sum_{m=1}^\infty {\Omega_T^{(m)}\over V}
 =
 i
 \sum_{n \ge 2, \sigma}
 {1\over S_{\{l_i^\sigma \}}}
 \int \prod_{i=1}^{n-1}d\Gamma_i^\sigma\,
 \langle \{l_i^\sigma \} |  
 \ln(1 + i{\cal T})
 | \{l_i^\sigma \} \rangle_\scrpt{conn}
 \;.
 \ee
 This is the expression given in Eq. (\ref{eq:omnor}) and, as we 
 remarked, it agrees with the result Norton \cite{Norton} obtained
 using a different approach. This in turn is equivalent to the expression
 obtained long ago by Dashen, Ma and Bernstein \cite{Dashen} from a
 non-relativistic analysis.


 %
 %
 \end{document}